\documentclass[10.75pt, a4paper]{article}
\usepackage{titlesec}
\titleformat{\section}
  {\bf\sffamily}
  {\thesection. }
  {5pt}
  {\MakeUppercase}
\renewcommand{\thesection}{\Roman{section}} 

\titleformat{\subsection}
  {\bf\sffamily}
  {\thesubsection. }
  {5pt}{}
  
\renewcommand{\thesubsection}{\Alph{subsection}}

\usepackage[affil-it]{authblk} 
\usepackage{etoolbox}
\usepackage{lmodern}

\usepackage{eurosym}

\usepackage[utf8]{inputenc}
\usepackage{geometry}
\usepackage[hidelinks]{hyperref}
\usepackage[citestyle=numeric,style=phys,biblabel=brackets,backend=bibtex,sorting=none, natbib]{biblatex}
\bibliography{bibliography}
\DefineBibliographyStrings{english}{andothers={\itshape et\addabbrvspace al\adddot}}

\usepackage{csquotes}
\usepackage[]{authblk} 
\usepackage{etoolbox}
\usepackage{lmodern}

\usepackage{amsmath}
\usepackage{enumerate}
\usepackage{enumitem}
\usepackage{graphicx}
\usepackage{siunitx}
\usepackage{float}
\usepackage[]{parskip} 

\makeatletter
\patchcmd{\@maketitle}{\LARGE \@title}{\fontsize{16}{19.2}\selectfont\@title}{}{}
\makeatother

\usepackage[normalem]{ulem}

\usepackage{graphicx}
\usepackage{dcolumn}
\usepackage{bm}
\usepackage{tikz}
\usetikzlibrary{math}
\usetikzlibrary{shapes.geometric, arrows}
\usetikzlibrary{positioning}
\usetikzlibrary{shapes,arrows}
\usetikzlibrary{intersections}
\usepackage{csvsimple}
\usepackage{changepage}
\usepackage[tbtags]{mathtools}
\usepackage{siunitx}
\usepackage{csquotes}

\usepackage{float}
\usepackage{subfigure}
\usepackage[
	nonumberlist, 				
	acronym,      				
	nomain,						
	nopostdot,					
]{glossaries}
\usepackage{glossary-superragged}
\usepackage[hidelinks]{hyperref}
\usepackage{color, colortbl}

\usepackage{wrapfig}

\usepackage{pgfplots}
\usepackage{tikz}
\usetikzlibrary{arrows}
\usepackage{amsmath}
\pgfplotsset{compat=newest}
\usepgfplotslibrary{fillbetween}
\usetikzlibrary{calc}
\def\centerarc[#1](#2)(#3:#4:#5)
{ \draw[#1] ($(#2)+({#5*cos(#3)},{#5*sin(#3)})$) arc (#3:#4:#5); }

\usepackage{amssymb}
\let\vec\mathbf
\usepackage{nicefrac}

\usepackage{abstract}

\usepackage{multirow}
\usepackage{tabularx}
\usepackage{booktabs}
\usepackage{array}
\usepackage{longtable}
\newcolumntype{L}[1]{>{\raggedright\let\newline\\\arraybackslash\hspace{0pt}}m{#1}}
\newcolumntype{C}[1]{>{\centering\let\newline\\\arraybackslash\hspace{0pt}}m{#1}}
\newcolumntype{R}[1]{>{\raggedleft\let\newline\\\arraybackslash\hspace{0pt}}m{#1}}

\newacronym{3d}{3D}{three dimensional}
\newacronym{am}{AM}{additive manufacturing}
\newacronym{fdm}{FDM}{fused deposition modeling}
\newacronym{ism}{ISM}{in-space manufacturing}
\newacronym{iss}{ISS}{International Space Station}
\newacronym{fcb}{FCB}{Functional Cargo Block}
\newacronym{dem}{DEM}{discrete element method}
\newacronym{md}{MD}{molecular dynamics}
\newacronym{dc}{DC}{direct-current}
\newacronym[plural=PFCs,firstplural=parabolic flight campaigns (PFCs)]{pfc}{PFC}{Parabolic Flight Campaign}
\newacronym{fft}{FFT}{Fast Fourrier Transform}
\newacronym{cad}{CAD}{Computer Assisted Design}
\newacronym{ptfe}{PTFE}{polytetrafluoroethylene}
\newacronym{ps}{PS}{polystyrene}
\newacronym{nasa}{NASA}{National Aeronautics and Space Administration}
\newacronym{esamm}{ESAMM}{Extended Structure Additive Manufacturing Machine}
\newacronym{amf}{AMF}{Additive Manufacturing Facility}
\newacronym{us}{US}{United States}
\newacronym{usa}{USA}{United States of America}
\newacronym{bmgs}{BMGs}{Bulk Metallic Glasses}
\newacronym{esa}{ESA}{European Space Agency}
\newacronym{si}{SI}{International System of Units, abbreviated from French \textit{Syst\`{e}me International (d'unit\'{e}s)}}
\newacronym{dlr}{DLR}{German Aerospace Center}
\newacronym{liggghts}{LIGGGHTS}{\acrshort{lammps} Improved for General Granular and Granular Heat Transfer Simulations}
\newacronym{lammps}{LAMMPS}{Large-scale Atomic/Molecular Massively Parallel Simulator}
\newacronym{sjkr}{SJKR}{Simplified Johnson-Kendall-Roberts}
\newacronym{ded}{DED}{Directed Energy Deposition}
\newacronym{slm}{SLM}{Selective Laser Melting}
\newacronym{sls}{SLS}{Selective Laser Sintering}
\newacronym{eva}{EVA}{Extra-Vehicular Activity}
\newacronym{sem}{SEM}{Scanning Electron Microscopy}
\newacronym{RPM}{RPM}{Ramdom Positioning Machine}
\newacronym{rpm}{rpm}{revolutions per minute}
\newacronym{rise}{RISE}{Research Internships in Science and Engineering}
\newacronym{daad}{DAAD}{German Academic Exchange Service, abbreviated from German \textit{Deutscher Akademischer Austauschdienst}}
\newacronym{fsm}{FSM}{finite-state machine}
\newacronym{ir}{IR}{infrared}
\newacronym{pcbs}{PCBs}{Printed Circuit Boards}
\newacronym{pcb}{PCB}{Printed Circuit Board}
\newacronym{mcr}{MCR}{Modular Compact Rheometer}
\newacronym{sff}{SFF}{Solid Freeform Fabrication}
\newacronym{uv}{UV}{ultraviolet}
\newacronym{abs}{ABS}{acrylonitrile butadiene styrene}
\newacronym{hpde}{HPDE}{high density polyethylene}
\newacronym{pei}{PEI}{polyetherimide}
\newacronym{bff}{BFF}{BioFabrication Facility}
\newacronym{lens}{LENS}{Laser Engineered Net Shaping}
\newacronym{cnc}{CNC}{Computer Numerical Control}
\newacronym{ebf3}{EBF$^3$}{Electron Beam Free-Form Fabrication}
\newacronym{leo}{LEO}{Low Earth Orbit}
\newacronym{pc}{PC}{polycarbonate}
\newacronym{crissp}{CRISSP}{Customisable Recyclable International Space Station Packaging}
\newacronym{Athena}{Athena}{Advanced Telescope for High-ENergy Astrophysics}
\newacronym{lbm}{LBM}{Laser Beam Melting}
\newacronym{bam}{BAM}{Federal Institute for Materials Research and Testing, abbreviated from German \textit{Bundesanstalt f\"{u}r Materialforschung und-pr\"{u}fung}}
\newacronym{pbf}{PBF}{powder bed fusion}
\newacronym{eb}{EB}{Electron Beam}
\newacronym{2d}{2D}{two dimensional}
\newacronym{4d}{4D}{four dimensional}
\newacronym{ft4}{FT4}{Freeman Technology 4 Powder Rheometer}
\newacronym{dsc}{DSC}{Differential Scanning Calorimetry}
\newacronym{pmma}{PMMA}{polymethylmethacrylate}
\newacronym{1g}{$1g$}{gravity on-ground}
\newacronym{mug}{$\mu g$}{microgravity}
\newacronym{bcm}{BCM}{Box Counting Method}
\newacronym{mct}{MCT}{Mode Coupling Theory}
\newacronym{gmct}{gMCT}{granular Mode Coupling Theory}
\newacronym{itt}{ITT}{Integration Through Transients}
\newacronym{mfc}{MFC}{Mass Flow Controller}
\newacronym{ct}{CT}{computed tomography}
\newacronym{xct}{XCT}{X-ray computed tomography}
\newacronym{cv}{CV}{curriculum vitae}
\newacronym{pi}{PI}{principal investigator}
\newacronym{osp}{OSP}{orthogonal superimposed perturbation}
\newacronym{npi}{NPI}{Network Partnering Initiative}
\newacronym{ecsat}{ECSAT}{European Centre for Space Applications and Telecommunications}
\newacronym{eac}{EAC}{European Astronaut Centre}
\newacronym{estec}{ESTEC}{European Space Research and Technology Centre}
\newacronym{fps}{fps}{frames per second}
\newacronym{pdf}{pdf}{probability density function}
\newacronym{al}{Al}{aluminium}
\newacronym{ss}{\textit{SS}}{\textit{Smooth Surface}}
\newacronym{rs}{\textit{RS}}{\textit{Rough Surface}}
\newacronym{rcp}{rcp}{random close packing}
\newacronym{iop}{IoP UvA}{Institute of Physics of the University of Amsterdam}
\newacronym{mp}{MP}{Institute of Material Physics for Space}
\newacronym{elgra}{ELGRA}{European Low Gravity Research Association}
\newacronym{zarm}{ZARM}{Center of Applied Space Technology and Microgravity}
\newacronym{piv}{PIV}{particle image velocimetry}
\usepackage[framemethod=tikz]{mdframed}
\usepackage{lipsum}
\usepackage{tcolorbox}
\usepackage[font=footnotesize,labelfont=bf]{caption}
\usepackage{sidecap}
\renewcommand{\figurename}{Fig.}

\usepackage{xcolor,hyperref}
\hypersetup{
   colorlinks,
   linkcolor={blue!50!black},
   citecolor={blue!50!black},
   urlcolor={blue!80!black}
} 

\title{Morphodynamics of chloroplast network control light-avoidance response\\
in the non-motile dinoflagellate \textit{Pyrocystis lunula}}
\author[1]{Nico Schramma\footnote{n.schramma@uva.nl}}
\author[1,2]{Gloria Casas Canales}
\author[1]{Maziyar Jalaal\footnote{m.jalaal@uva.nl}}

\affil[1]{Van der Waals-Zeeman Institute, University of Amsterdam}
\affil[2]{Institute for Biodiversity and Ecosystem Dynamics, University of Amsterdam}

\begin{document}

\begingroup
\sffamily
\maketitle
\endgroup

\begin{abstract}
Photosynthetic algae play a significant role in oceanic carbon capture. Their performance, however, is constantly challenged by fluctuations in environmental light conditions.
Here, we show that the non-motile single-celled marine dinoflagellate \textit{Pyrocystis lunula} can internally contract its chloroplast network in response to light.
By exposing the cell to various physiological light conditions and applying temporal illumination sequences, we find that network morphodynamics follows simple rules, as established in a mathematical model.
Our analysis of the chloroplast structure reveals that its unusual reticulated morphology constitutes properties similar to auxetic metamaterials,
facilitating drastic deformations for light-avoidance, while confined by the cell wall. 
Our study shows how the topologically complex network of chloroplasts is crucial in supporting the dinoflagellate's adaptation to varying light conditions, thereby facilitating essential life-sustaining processes.

\end{abstract}

\vspace{8pt}
Light is fundamental for life, yet excessive exposure can be detrimental for biological processes. Photosynthetic organisms have developed multiple strategies across scales to cope with fluctuations of light, from molecular adaptation responses such as Non-Photochemical Quenching (NPQ) to the biased growth towards light (phototropism) on organismal scale~\cite{Ruban2015,Li2009, Liscum2014}. Photoadaptation in the form of increased thermal dissipation of photo-excited chlorophyll, which displays a major component of NPQ, occurs within tens of seconds to minutes~\cite{Krause1991}, while plant tropism takes longer time scales of hours to days~\cite{Darwin1880,Liscum2014}. Another photoadaptation mechanism, at a single-cell level, includes the motion and rearrangement of chloroplasts to optimize light absorption or to avoid photo-damage~\cite{Senn1908,Kasahara2002,Park1996,Li2009,Wada2013, Schramma2023}. 
Such strategies are fundamental for survival in fluctuating environments,
displaying interesting features such as computation and integration in the context of plant phototropism~\cite{Meroz2019,Moulton2020,Riviere2023}, and dynamical phase transitions of chloroplast motion~\cite{Schramma2023}.
In land plants, the collective motion of individually sensing and moving disk-shaped chloroplasts~\cite{Wada2013,Wada2018} leads to large-scale re-arrangements inside leaf cells, aiming to optimize for light uptake while avoiding strong light.
However, a vast amount of photosynthesis happens in aquatic environments, especially the oceans
~\cite{Field1998,Behrenfeld2014,Cavicchioli2019}. 
Some single-celled algae can actively swim away from or towards light~\cite{Jekely2009,Drescher2010,Jin2020,DeMaleprade2020} and even less complex organisms such as cyanobacteria, like \textit{Trichodesmium}~\cite{Pfreundt2023} and \textit{Synechocystis}~\cite{Schuergers2016,Bhaya2004}, can move collectively or individually as a response to light.
Non-motile photosynthetic organisms, nonetheless, have adopted different strategies by moving their chloroplasts within their cell bodies~\cite{Haupt1982,Williamson1993}. \\
Here, we study the light adaptation strategy of a non-motile marine dinoflagellate.
Dinoflagellates are a large and distinct group of photosynthetic and mixotrophic algae~\cite{Lin2011,Cohen2021} that, unlike most other photosynthetic organisms, underwent tertiary endosymbiosis by engulfing algae containing a secondary plastid~\cite{Sagan1967, Gray1992, Gray2017, Bhattacharya2004}. This process led to chloroplasts with three membranes~\cite{Fast2001}. 
 Mostly known for their bioluminescence~\cite{Biggley1969,Tesson2015,Jalaal2020}, the non-motile, $\sim 50-100\,\mathrm{\mu m}$-sized dinoflagellate \textit{Pyrocystis lunula}, native to warm waters worldwide, can reorganize its internal architecture to switch from a photosynthetic day-phase to a bioluminescent night-phase. This reorganization is orchestrated by actively moving their chloroplasts and bioluminescent organelles (scintillons) towards and away from the cell center, following a circadian rhythm~\cite{Topperwien1980,Fritz2000,Heimann2009}. 
Interestingly, it has been observed that the same intracellular motion can be triggered by strong light~\cite{Swift1967}, leading to a compression of the chloroplast within just a couple of minutes, without deformation of the thick cell wall~\cite{Swift1970}.
This observation raises the question of how this drastic intracellular rearrangement can be coordinated under confinement and within such a short time frame.
Here, we study the dynamics of chloroplast motion as a light adaptation mechanism in \textit{P. lunula} and report that it features an active reticulated chloroplast network, capable of fast morphological changes. This enables the organelle to undergo significant deformations for efficient photo-avoidance motion in response to environmental changes.


\begin{figure}[h]
    \centering
    \includegraphics[width=1\textwidth]{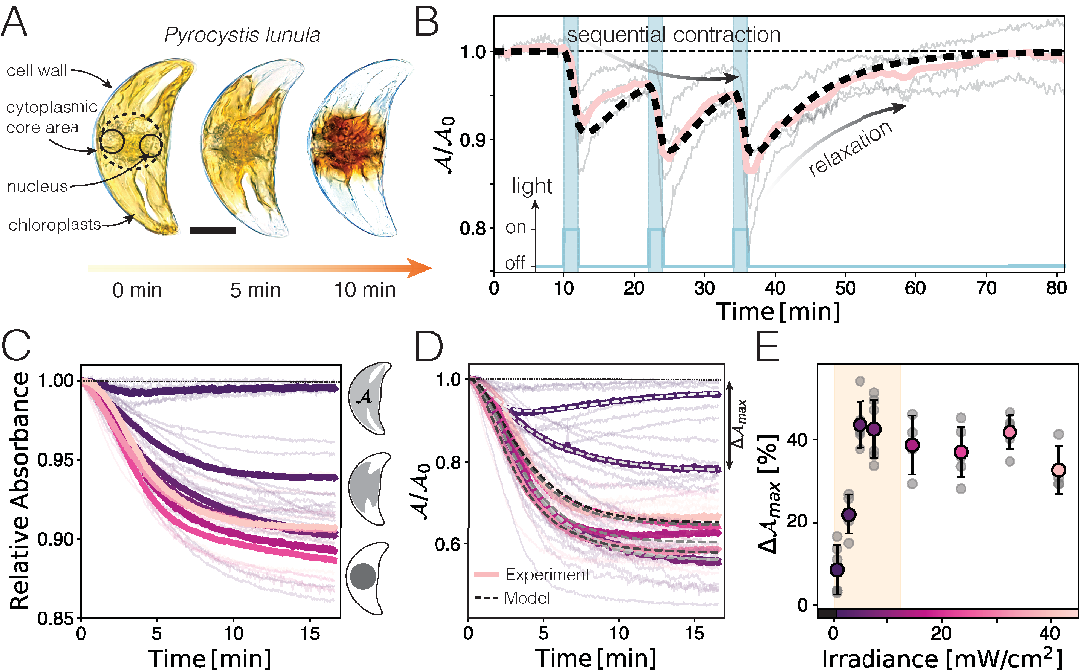}
    \caption{\textbf{Chloroplast area of \textit{Pyrocystis lunula} undergoes rapid changes under strong light illumination.} (\textbf{A}) Upon light activation, the chloroplast (yellow) contracts and compacts (darker color), achieving significant change in less than $10\,\mathrm{min}$. Scalebar: $20\,\mathrm{\mu m}$. (\textbf{B}) Periodic light stimulation (blue curve), causes the chloroplast area $\mathcal{A}$ (normalized by initial area $\mathcal{A}_0$, $N=4$, orange curve: mean) to decrease under strong light and expand under dim light. The dynamics are well described by an active viscoelastic model (dashed line).  (\textbf{C} and \textbf{D}) Relative absorbance and area decline in response to varying light intensities in specimens initially adapted to dim light, indicating increased light transmission in the light-avoidant state. Higher light intensities cause a stronger response (colorscale in \textbf{E}). (\textbf{E}) The maximum area reduction correlates with light irradiance. Beyond physiologically tractable light conditions (yellow box) the contraction response saturates to a maximal decrease of the chloroplast area by $40\,\%$. }
    \label{fig:Figure1}
\end{figure}

\subsubsection*{Cytoplasmic space contracts under strong light}
We studied the adaptation of the chloroplast area of individual \textit{Pyrocystis lunula} cells to white-light exposure under physiologically relevant light conditions of its natural habitat at depth of $60-100\,\mathrm{m}$~\cite{Bhovichitra1977} (Supplementary Text Section I). Upon white-light stimulation, chloroplasts move towards the cytoplasmic core area within approximately $10\,\mathrm{min}$ (Fig.~\ref{fig:Figure1}A)~\cite{Swift1967}. 
During this process, the relative absorbed light (Fig.~\ref{fig:Figure1}C) and the projected area $\mathcal{A}$ of the chloroplast (Fig.~\ref{fig:Figure1}D) exhibit similar dynamics for various light irradiances ranging from $2.8$ to $41.4\,\mathrm{mW/cm^2}$ (fig.~\ref{fig:FigureA1}, Movie S1). After a short time of approximately $1-2\,\mathrm{min}$, characterized by a positive curvature, both the projected area and the absorbed light follow an exponential decay, eventually reaching saturation at levels dependent on the light intensity. These measurements are correlated: as the chloroplast retracts, more light can pass through the cell, leading to reduced absorption of up to $10\,\%$.\\
Contrary to the contraction scenario described above, exposure to dim white light ($0.6\,\mathrm{mW/cm^2}$) triggers a transient response wherein the chloroplast initially contracts towards the cell center but eventually expands again (Fig.~\ref{fig:Figure1}D). 
These observations share similarities with terrestrial plants' transient photo-response at the intermediate light intensity~\cite{Luesse2010,Labuz2022}.
Large stimulation-irradiances, which exceed both the estimated ecologically relevant light conditions of $0.2-12\,\mathrm{mW/cm^2}$ (Supplementary Text Section I) and the culturing conditions $I=0.27 \,\mathrm{mW/cm^2}$ (Materials and Methods) lead to maximal contraction of $\Delta \mathcal{A}_{max} \approx 40\,\%$, corresponding to the size of the cytoplasmic core area (Fig.~\ref{fig:Figure1}A,E).

\subsubsection*{Dynamical testing and mathematical model of chloroplast contraction}
Alternating white- (on) and red-light (off) irradiation controls the chloroplast contraction and relaxation dynamics towards and away from the cytoplasmic core area (Fig.~\ref{fig:Figure1}B, Movie S2,S3): the chloroplast network rapidly responds to white-light and slowly relaxes upon red-light exposure. Long-time illumination with weak red light ($0.2\,\mathrm{mW/cm^2}$) leads to a complete expansion of the chloroplasts over the entire cell (Fig.~\ref{fig:Figure1}B, Time$>50\,\mathrm{min}$).\\
The dynamical tests verify a mathematical model (dotted lines Fig.~\ref{fig:Figure1}B,D, \ref{fig:Figure3}G, \ref{fig:Figure4}C,D) (Materials and Methods, Supplementary Text Section II), based on two chemical species that control a light-avoidant and a light-accumulation response (contraction or expansion), respectively (fig.~\ref{fig:FigureA2}).\\
The model enables the extraction of two relevant chemical signaling timescales, $\tau_1$ and $\tau_2$, along with a timescale for active contraction $\tau_{KV}$ and expansion $\tau_{KV}^*$. 
%
We find that the chemical signaling of a photo-avoidance response occurs at a time scale of $\tau_1=1.8\,\pm0.5\,\mathrm{min}$ (mean $\pm$ SD, $N=44$), while a photo-attractive signal leading to a transient response is noted at $\tau_2 = 2.5\,\pm 0.9\,\mathrm{min}$ ($N=6$). 
The timescales $\tau_{KV}$ and $\tau_{KV}^*$ emerge from the dynamics of the driving mechanism, relying on actin, microtubules, and molecular motors such as myosin~\cite{Heimann2009} (also see Supplementary Text Section III for inhibitory treatment), and found to be $\tau_{KV} = 2.7\pm0.9\,\mathrm{min}$ ($N=44$) for contraction and $\tau_{KV}^*=6\pm3\,\mathrm{min}$ ($N=6$) for expansion of the chloroplast. Notably, our model can be interpreted as a filter for environmental stimuli (Supplementary Text Section II), effectively filtering out high-frequency noise (with short time scales) such as the light fluctuations of surface waves (Supplementary Text Section I).

\begin{figure}[t!]
    \centering
    \includegraphics[width=1\textwidth]{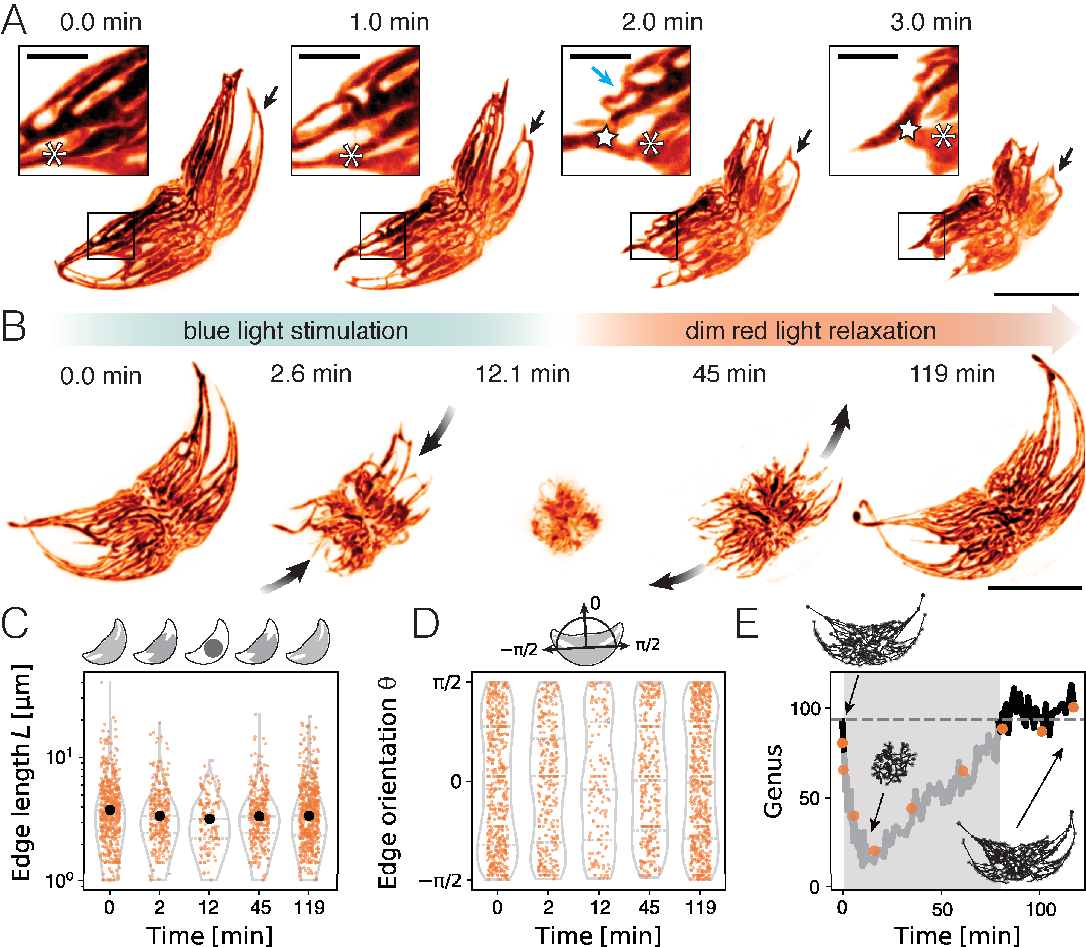}
    \caption{\textbf{Dynamic photo-avoidance and photo-accumulation of the chloroplast reticulum.} Autofluorescence imaging of chloroplasts reveals reticulated structure. (\textbf{A}) Light stimulation beginning at $t=0\,\mathrm{min}$ causes the chloroplast network to contract, by moving strands towards the cells' center and deforming them (arrows). A compact structure is achieved by reducing hole sizes of the network. Scale bar: $40\,\mathrm{\mu m}$. Inset: Examining an area of dynamical deformation of a strand of the chloroplast network, two nodes (asterisk and star) change their position as well as their distance from one another. The blue arrow indicates a deforming cytoplasmic strand leading to the closure of a hole in the network. Scale bar: $10\,\mathrm{\mu m}$. (\textbf{B}) Chloroplast contraction under blue light stimulation and subsequent expansion under the dim red light environment. (\textbf{C}) Edge length statistics of the chloroplast network (from B) over time. Large edges shorten and extend again. (\textbf{D}) Distribution of orientation of edges. Orientations align with the major axis of the cell, but during contraction, the distribution flattens. (\textbf{E}) Topological measurements over time, indicating similarity of structures before and after the stimulation. Black line: Genus estimated by Betti number $\beta_1$ of the network. Orange points: Genus measured from Euler characteristic $\chi$ of the mask image. During contraction chloroplast strands touch and seemingly close holes leading to an apparent decrease in detected genus (gray area). Genus before contraction and after expansion has similar values. }
    \label{fig:Figure2}
\end{figure}

\subsubsection*{Topologically complex chloroplast network allows efficient contraction}

The contraction of the projected area reaches up to $40\,\%$ of the cell area, necessitating a remarkably large deformation of the cytoplasmic material.
Such deformations within hard confinement, in this case, constituted by the cell wall of \textit{P. lunula}~\cite{Swift1970}, are subject to physical constraints. To elucidate these constraints, we draw analogies to the compression of fluids and solids:
an incompressible fluid cannot ``contract'' uniformly towards the center under confinement, as its volume must be conserved. Similarly, an elastic solid, when compressed from one direction, will expand in orthogonal directions, a behavior characterized by a positive Poisson ratio ($\nu=0.5$ for isotropic incompressible elastic materials in three dimensions). Under confinement, this property ($\nu > 0$) leads to effective strain-stiffening, as the expansion into orthogonal directions is prevented and therefore internal stresses redirect. 
\begin{figure}[t!]
    \centering
    \includegraphics[width=1\textwidth]{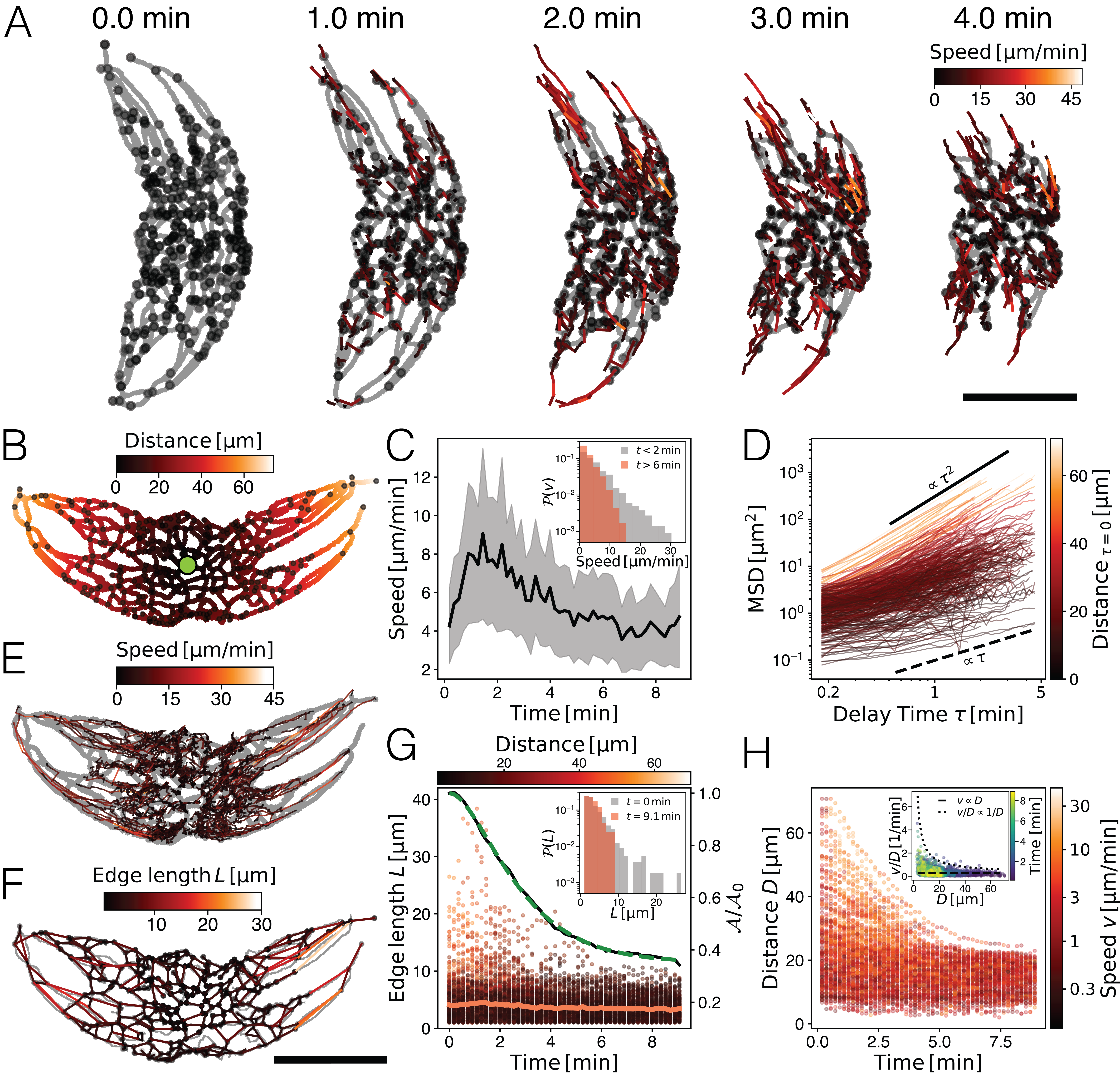}
    \caption{\textbf{Network dynamics of the chloroplast reticulum}. (\textbf{A}) Time series of chloroplast contraction dynamics. The nodes (black dots) of the underlying three-dimensional network move on paths inward (color: speed). Gray: 2d projection of the underlying skeleton.
     (\textbf{B}) Distance-map from the center point (green dot) on a 2d-projected mask of the experiment, as in (A) at $t=0\,\mathrm{min}$.(\textbf{C}) Average individual node speed (black line: mean $\pm$ SD) increases during contraction and subsequently decreases. Inset: the speed distribution is heavy-tailed. Initially fast trajectories ($t<2\,\mathrm{min}$) slow down at long times $t>6\,\mathrm{min}$.
    (\textbf{D}) Mean-squared displacement (MSD) of individual trajectories ranges from diffusive trajectories ($\mathrm{MSD}\propto \tau$) in the center (colors corresponding to the distance from the center (B) at the beginning of the trajectory) to purely ballistic trajectories ($\mathrm{MSD}\propto \tau^2$) originating at the periphery. (\textbf{E}) Trajectories are mapped on the skeleton of the mask (gray) and colored according to their local speed reaching up to $30\,\mathrm{\mu m/min}$.  (\textbf{F}) Network representation of the chloroplast. Edges are colored by length. The network closely represents the skeleton of the mask (gray). (\textbf{G}) The distribution of edge lengths of the network decays over time. The mean edge length (orange line) is hardly affected. Colorbar: edge distance to the center corresponding colors in (B) and (D). The shrinkage of the longest edges follows a similar trend to the decrease in the relative projected area of the chloroplast network (black line, right axis). The model fit to the projected area curve (dashed green line) shows good agreement with the experimental observations. Inset: The network has initially ($t=0\,\mathrm{min}$) a few very long trajectories (grey), which disappear at long times (red).  (\textbf{H}) Correlation of distance and speed. Far-distanced nodes travel at higher speeds (brighter colors) as compared to centrally located nodes. Note the logarithmic colorbar. Inset: ``Strain rate'' (speed over distance) of every node is constant at large distances. The solid line corresponds to $v\propto D$ which is expected for elastic deformation at a constant strain rate. The dotted line corresponds to constant $v$. Colormap represents time. All scale bars: $40\,\mathrm{\mu m}$}.
    \label{fig:Figure3}
\end{figure}
However, structured metamaterials defy positive Poisson ratios by allowing non-linear deformations such as buckling or the action of rotating hinges~\cite{Florijn2014,Bertoldi2017}. Auxetic behavior, characterized by $\nu< 0$, naturally arises in polymer networks~\cite{Domaschke2019}, foams~\cite{Lakes1987}, and poro-elastic materials such as cork ($\nu\approx 0$)~\cite{fortes1989poison}, enabling uni- or multi-directional contraction, and thus facilitating efficient deformations even within confined spaces.\\
Using confocal auto-fluorescence imaging of the chloroplasts (Materials and Methods) we uncover the chloroplast reticulum  (Fig.~\ref{fig:Figure2}A). This intricate network structure shows similarities to observations made in the context of spontaneous diurnal chloroplast relocations \cite{Fritz2000,Heimann2009} and rapid changes of buoyancy in the related species \textit{Pyrocystis noctiluca} \cite{Larson2022}. 
%
Continuous blue light stimulation ($\lambda=470\pm50\,\mathrm{nm}$) of the cell, induces contraction of the chloroplast network over time.
We uncover two mechanisms which choreograph this chloroplast photo-adaptaion motion: cytoplasmic strands move towards the center while they simultaneously contract in a manner reminiscent of buckling (Fig.~\ref{fig:Figure2}A, Movie S4-S6).
Buckling, or inward-folding, allows the structure to compact into the space between the cytoplasmic strands ($\nu\lesssim 0$), circumventing strain-stiffening typically expected from uniform bulk contractions under confinement ($\nu>0$). Moreover, we observe a notable thickening of some strands during contraction, indicating a flow of material within the structure (Fig.~\ref{fig:Figure2}A, Movie S4-S6).  \\
%
Under dim red light, the chloroplast expands again over a longer time scale (Fig.~\ref{fig:Figure2}B, Movie S6). Although the chloroplast network appears slightly different before and after the contraction and expansion, the main geometrical features remain unchanged: when fully spread, the network is characterized by a few very long cytoplasmic strands (Fig.~\ref{fig:Figure2}C), which extend outwards along the cell's long body axis (Fig.~\ref{fig:Figure2}D). 
During contraction, these distinctive features are lost as the chloroplast reticulum obtains a more spherical shape.
Remarkably, despite these morphological changes, the network topology remains largely unchanged before contraction and after expansion, suggesting a permanent connection among nodes without dynamic rewiring of the network (Fig.~\ref{fig:Figure2}E). The apparent loss of holes, as indicated by the decline in Genus, can be attributed to the increased contact between strands, which complicates the identification of individual strands.\\
%
We analyze the spatiotemporal dynamics of the networks' nodes over time (Fig.~\ref{fig:Figure3}A,B,E, Materials and Methods, Movie S7) to quantify the 
contraction process. Our findings reveal a wide distribution in node speeds, which vary over time (Fig.~\ref{fig:Figure3}A,C). Predominantly, nodes close to the cytoplasmic core area exhibit slow, diffusive motion characterized by a time-linear mean-squared displacement ($\mathrm{MSD}\propto \tau$). Conversely, nodes located farther from the cell center move ballistically ($\mathrm{MSD}\propto \tau^2$) (Fig.~\ref{fig:Figure3}D).
Thus, longer strands, initially located far from the cytoplasmic core area, move rapidly towards the center (Fig.~\ref{fig:Figure3}E,H), while reducing their length, resulting in the reduction of chloroplast's 2D-projected area $\mathcal{A}$ (Fig.~\ref{fig:Figure3}G). 
Interestingly, the strands follow along their initial configuration, as indicated by the overlay of trajectories on the network's 2D projection at $t=0\,\mathrm{min}$ in figure~\ref{fig:Figure3}A,E.  This may reflect a signature of their actin- and microtubule-mediated driving mechanism~\cite{Heimann2009} and a co-alignment to those networks.
Further analysis of the network, demonstrates that the speed of the nodes scales with their distance from the center, $v \propto D$, aligning with the expected behavior from the one-dimensional contraction of an elastic material under a constant strain rate (Fig.~\ref{fig:Figure3}H, inset). This observation confirms that the complex topology of the network enables the material to contract, dominantly, along one dimension in a manner similar to an elastic material ($\nu\sim 0$). These findings have been valuable in developing the mathematical model (Supplementary Text Section II).
\subsubsection*{Chloroplast contraction shows local and global responses to local stimulation}

To further shine light on the sensing mechanism, we locally stimulate the cell (Materials and Methods) with a blue low-power laser ($\lambda=488\,\mathrm{nm}$, $3.25\,\mathrm{mW/cm^2}$). Although this power level 
may seem weak compared to the full-spectrum irradiance of white-light stimulation used in other experiments, it closely matches the optimal absorption peak of phototropins~\cite{Briggs2002}. Phototropins, crucial photosensors for chloroplast positioning in terrestrial plants~\cite{Suetsugu2007, Harada2007,Wada2013}, are also present in \textit{Pyrocystis lunula}~\cite{Menghini2021}, potentially accounting for the observed strong response.\\
\begin{figure}[t!]
    \centering
    \includegraphics[width=1\textwidth]{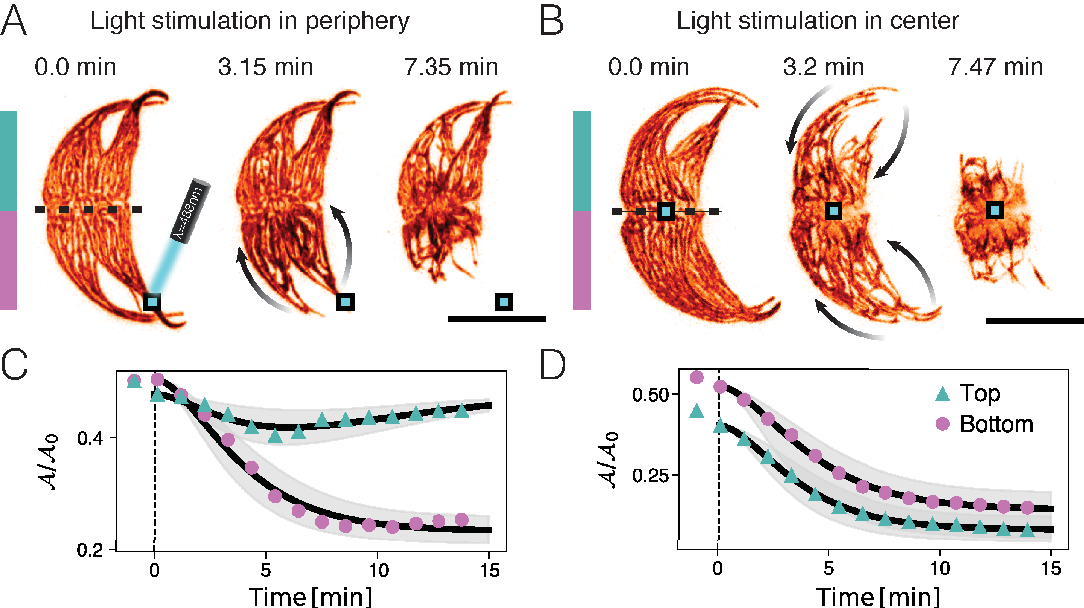}
    \caption{\textbf{Local sensing leads to local or global response} (\textbf{A}) Local stimulation in the periphery with blue light ($\lambda=488\,\mathrm{nm}$, $I=3.25\,\mathrm{mW/cm^2}$) leads to a one-sided response away from the stimulation region within a few minutes. Note the contraction of the network in the opposite site. Blue square indicates stimulation region. (\textbf{B})  Light stimulus in the center (blue square) leads to the global movement of the chloroplast network towards the center (arrows). (\textbf{C}) Area contraction in (A) upper and lower half of the cell (divided by dashed line in (A)). The lower part contracts rapidly, while the upper half undergoes a transient response. (\textbf{D}) Area contraction in (B). Upper and lower half of the cell undergo a symmetric response. See Movies S8-S10. Scale bar: $40\,\mathrm{\mu m}$.}
    \label{fig:Figure4}
\end{figure}
We illuminate a $6.8\times6.8\,\mathrm{\mu m^2}$ region at both the center and periphery of the algae, respectively (Fig.~\ref{fig:Figure4}A,B, Movies S8-S10). Peripheral stimulation leads to a localized contraction of one side of the chloroplast network towards the cytoplasmic core area. The opposite side of the chloroplast network moves less pronounced, but within the temporal resolution limit of $20-40\,\mathrm{s}$.
This rapid onset suggests that a fast diffusing signal triggers a long-range response across the cell of $L\approx80-100\,\mathrm{\mu m}$ within $\tau\approx20-40\,\mathrm{s}$ after light-stimulation begins, implying a diffusion coefficient on the order of $D\approx L^2/\tau \approx160-500\,\mathrm{\mu m^2/s}$. 
Interestingly, the chloroplast retraction on the not-stimulated region is transient or less pronounced, as observed in small light intensity stimulation (Fig.~\ref{fig:Figure1}D, Movie S8,S9), suggesting that the transmitted signal is depleted, as expected for diffusive signaling.\\
Surprisingly, upon central stimulation in the cytoplasmic core area, all chloroplast strands contract towards the cell center, within similar time scales (Fig.~\ref{fig:Figure1}B,D, Fig.~\ref{fig:Figure3}A,G). This outcome reveals a symmetry-broken response: under peripheral stimulation, the chloroplast network contracts to move \textit{away} from the light source, whereas central stimulation induces movement \textit{towards} it. This pattern suggests that the chloroplast network's contraction mechanism is inherently directed inward in response to strong light, regardless of the signal's location.

\subsubsection*{\large Discussion}
In our experiments, we show that the chloroplasts of \textit{Pyrocystis lunula} retract towards the cell's center under strong white or blue light conditions and expand under weak red light conditions. This bidirectional movement of chloroplasts towards and away from light, mirrors the chloroplast photo-relocation motion seen in leaves of green plants~\cite{Senn1908,Li2009,Wada2013,Liscum2014}, nonetheless, using a fundamentally different dynamics. 
The significant light-induced retraction of chloroplasts leads to increased light transmission through the cells (Fig.~\ref{fig:Figure1}C), suggesting that, similar to green plants, \textit{Pyrocystis lunula} employs this mechanism as a means of light avoidance~\cite{Senn1908,Kasahara2002,Li2009,Wada2013}.
Notably, under low white light conditions, we observe a transient response in the chloroplasts -- initial fast contraction followed by slow expansion -- suggesting a timescale-separated competition between these processes.
This shows similarities with the observed counteraction of phototropin 1 and 2-mediated chloroplast motion~\cite{Luesse2010,Banas2012}.
In fact, the transcriptome of \textit{P. lunula} bears various phototropin 1 and phototropin 2-like sequences and LOV domains~\cite{Menghini2021}, pointing towards potential similarities in light sensation. The similarity of this organism's light response to green plants is surprising, as the origin of chloroplasts in dinoflagellates is very distinct~\cite{Lin2011,Bhattacharya2004}.\\
At light intensities exceeding the natural physiological conditions of dinoflagellates,
chloroplasts contract fully towards the cell center. Under such extreme conditions, the crowding of the chloroplast strands poses a mechanical limit for contraction, and consequently for this photo-avoidance mechanism. 
However, within their native physiological light conditions, \textit{Pyrocystis lunula} responds via light-dependent chloroplast compaction, indicating a gradual adaptation response to various light conditions.
Furthermore, we find that the relaxation phase following strong light-induced contraction occurs over a longer time scale, suggesting a different driving mechanism for the expansion of the chloroplasts.\\
The large-scale transport of organelles observed likely depends on the coordinated action of the actin and microtubule networks, together with molecular motors such as myosin, as has been demonstrated in the context of the diurnal intracellular reorganization between the photosynthetic phase during the day and bioluminescent phase at night~\cite{Heimann2009}, in line with pharmacological perturbation experiments for light-adaptation (Supplementary Text III, fig.~\ref{fig:Figure_Pharma}) in which we confirm that actin is necessary for bi-directional chloroplast motion.
However, the exact driving mechanism of the chloroplast relocation in \textit{P. lunula} remains unidentified and may differ from that of green plants, where chloroplast movement is primarily driven by the assembly of short actin (cp-actin) filaments and the transmission of polymerization forces towards the plasma membrane~\cite{Wada2013, Wada2018,Banas2012}.\\
We elucidate the details of the reticulated morphology of chloroplasts and show that such a structural ``design'' offers mechanical advantages. Structured metamaterials, like this chloroplast morphology, facilitate buckling and other complex deformations~\cite{Florijn2014,Bertoldi2017}, enabling efficient chloroplast contraction under the confinement of the cell wall.
%
\textit{Pyrocystis noctiluca}, another species within the \textit{Pyrocystis} genus, was found to have a reticulated cytoplasm, assisting with the
vertical migration~\cite{Larson2022}, showing another intricate link between the morphology of cytoplasmic space and its function in dinoflagellates.\\ 
We also showed that the dynamics of chloroplast contraction can be effectively modeled using a ``visco-elastic'' framework with chemically controlled stress applications. 
Although our coarse-grained model does not pinpoint the precise origins of observed elasticity and viscosity -- whether from passive or active cellular components -- it importantly enabled us to identify adaptation time scales and compare them with the ecologically relevant fluctuations.
These time scales suggest that chloroplast motion serves as a feasible light adaptation strategy for environmental light variations persisting longer than $3-5\,\mathrm{min}$. Indeed, such fluctuations notably exceed the duration of second-long light changes induced by waves but are in line with the motion of clouds obscuring the sun \cite{Stramska1998} and might complement NPQ.\\
%
Our experiments have also shown that chloroplast contraction is driven by local sensing, with directed relocation observed when chloroplasts are locally stimulated (Fig.~\ref{fig:Figure4}). However, even though locally stimulated, the chloroplast network on the opposite side of the cell contract within $20-40\,\mathrm{s}$, indicating a long-ranged signal transfer via fast diffusive signals ($D\approx 160-500\,\mathrm{\mu m^2/s}$), such as calcium, recognized for its important roles in photosensory downstream signaling~\cite{Harada2007} and in \textit{P. lunula} bioluminescence~\cite{Jin2013}. 
Interestingly, stimulation in the cytoplasmic core prompts chloroplasts to move towards rather than away from the center. This counter-intuitive behavior might result from the network's topologically conserved structure and a photo movement that is inherently biased towards the cell center. This hypothesis, however, needs further investigation. 
Moreover, the intricate relationship between the chloroplast and nucleus, with the latter hosting a significant portion of the chloroplast genome~\cite{Surpin2002,Li2009}, suggests that chloroplast movement towards the nucleus in the cytoplasmic core~\cite{Fritz2000} could also serve a photoprotective function, shielding genetic material from intense light damage. 
Our study provides the first evidence for such a mechanism in dinoflagellates, however, a comprehensive examination of the chloroplast-nucleus relationship is necessary to fully understand these dynamics.\\ 
Overall, the complex relationship between the geometry and topology of chloroplast structure and its dynamics provides a fertile ground for exploring intriguing physical dynamics with significant physiological implications, in the context of light-life interactions.




\paragraph{Acknowledgements}
We thank Ronald Breedijk for help with confocal imaging and localized light stimulation.
We thank Joachim Goedhart, Mark Hink, Yuri Z. Sinzato, Jonas Veenstra, Corentin Coulais and Friedrich Kleiner for fruitful discussions. 
Confocal imaging was supported by the van Leeuwenhoek Centre for Advanced Microscopy, Section Molecular Cytology, Swammerdam Institute for Life Sciences, University of Amsterdam, a EuroBioImaging-node.
MJ acknowledges the ERC grant no.~"2023-StG-101117025, FluMAB".

\paragraph{Author Contribution}
Nico Schramma: Conceptualization (lead); Investigation (equal); Validation (equal); Writing – original draft (lead); Visualization (lead); Formal analysis (lead); Writing – review and editing (equal); Software (lead); Methodology (lead) \\Gloria Casas Canales: Conceptualization (supporting); Investigation (equal); Validation (equal) ;Writing – original draft (supporting); Visualization (supporting); Writing – review and editing (equal) \\Maziyar Jalaal: Conceptualization (supporting); Writing – original draft (supporting); Visualization (supporting); Writing – review and editing (equal); Methodology (supporting);  Supervision (lead); Funding (lead)
\paragraph{Competing interests}
The authors declare no competing interests.
\paragraph{Data and materials availability}
Data is available upon request at the corresponding author.

\clearpage

\section*{\huge Appendix}
\section*{\large Materials and Methods}
\subsubsection*{Cell culture}

\textit{Pyrocystis lunula} (Schütt) is cultured in $f/2$ medium in an incubator (Memmert) at $20\,\mathrm{^\circ C}$ and a $12:12$ day-night cycle with light irradiance at $0.27\,\mathrm{mW/cm^2}$.


\subsubsection*{Brightfield microscopy}
We perform bright-field microscopy with a Nikon TI2 E microscope. Images are acquired using a Photometrics BSI Express camera at a frame rate of $1- 2\,\mathrm{frame/s}$. We use a $40x/1.2\mathrm{NA}$ PLAN Apochromat objective and simultaneously measure $3-5$ different positions using a motorized xy-stage totaling $4-8$ cells for every light intensity setting ($0.6, 2.8, 4.9, 7.6, 14.5, 23.4, 32.4, 41.4 \,\mathrm{mW/cm^2}$). To allow for chloroplast expansion, imaging was performed using a red filter with a cut-on wavelength of $\lambda_c = 625\,\mathrm{nm}$, at light intensities of $0.1 - 0.4\,\mathrm{mW/cm^2}$. 
Periodic white ($7.6\,\mathrm{mW/cm^2}$) and red light ($0.1\,\mathrm{mW/cm^2}$) stimulation was controlled through a custom script in microManager \cite{edelstein2014}.

\subsubsection*{Confocal microscopy}
Global simulation experiments are performed with Nikon Ti Eclipse microscope equipped with a $60\times$/1.49NA or $40\times$/1.3NA oil objective, a confocal spinning disk unit (Yokogawa CSU-X1) with a microlense array, and an Andor iXonEM+ 897 electron-multiplying charged-coupled device (EMCCD) camera. Chlorophyll autofluorescence is stimulated at $640\,\mathrm{nm}$ wavelength while the emission bandpass filter ($680-740\,\mathrm{nm}$) is used. 
By using a piezo z-stage we record $50-120$ z-steps (step size $0.3-1\,\mathrm{\mu m}$) within $11-20,\mathrm{s}$ with a $30-60\,\mathrm{ms}$ excitation time.
The brightfield path of the microscope was equipped with a $470\pm50\,\mathrm{nm}$ blue filter to stimulate the cell at an intensity of $0.6\,\mathrm{mW/cm^2}$. 

\subsubsection*{Local stimulation}
Local stimulation experiments are performed using a $63\times$ PLAN APO IR objective on a Nikon TI body with Nikon A1 resonant scanner. Imaging is performed using a $637\mathrm{nm}$ laser line and emission filter at $650\,\mathrm{nm}$.
A whole z-stack consists of $80$-$100$ steps of $0.5\,\mathrm{\mu m}$ size. The pinhole diameter is fixed to $1.2$ Airy disk diameters.
Local stimulation is performed by applying a $488\,\mathrm{nm}$ laser at low irradiance $3.25\,\mathrm{mW/cm^2}$ in a small scanning region of $6.8\times 6.8\,\mathrm{\mu m^2}$. One z-stack takes between $45\,\mathrm{s}$ and $60\,\mathrm{s}$ and the stimulation time is $3.9\,\mathrm{s}$ for center-stimuli or $7.8\,\mathrm{s}$ for peripheral stimuli. 

\subsection*{Image processing}

Image analysis was performed in Fiji \cite{FIJI} and Python using Napari \cite{Napari} and scikit-image \cite{van2014scikit}. In the following paragraphs we will outline the different image processing steps for the different data we acquired.

\paragraph{Measuring chloroplast area from brightfield data}
We customized a Fiji macro, which measures brightfield intensity in a hand-annotated region of interest (ROI) for sequential stimulation experiments, or generated ROI for constant illumination experiments. The ROI is generated by Gaussian smoothing of the first frame with a $21px$ kernel and subsequent thresholding using a triangle filter. Morphological closure with a $51px$-diameter disk-shaped mask helps filling holes within the mask.  
To measure the area we calculate a the mean value of the background, subtract it by one standard deviation and employ it as a threshold value to discriminate the chloroplast material from the background light transmitted through the organism. 

\paragraph{Analyzing 3D confocal data}
We segment and skeletonize 4d stacks (x,y,z,t) by using a script developed within Napari. 
The following steps are performed for all time steps.
First, we estimate the loss of light intensity deeper within the sample by fitting the Beer-Lambert law $I(z)=I_0e^{-z/\lambda}$ and correcting the z-stack accordingly.
Next, we blur the x-y plane with a $1\,\mathrm{px}$-wide Gaussian and subsequently use a top-hat filter with a $(2.8,2.8,1)\,\mathrm{\mu m}$ kernel. Then, we use the triangle method to threshold and perform binary closing with a $(1,1,1)\,\mathrm{\mu m}$ kernel.
We label the obtained mask with a connected-component method and reject small labels $<20000\,\mathrm{px}$.
Three-dimensional graph analysis was performed after skeletonizing the label image \cite{lee1994building} and generating a NetworkX graph \cite{NetworkX}. Some closely located nodes of the graph are merging over time, hence we coarse-grain nodes that are less than $5px$ away from each other by deleting them and setting one node in the middle between them. Similarly, we reject end-nodes that have a single edge less than $9px$ long. 
The nodes of the graph represent the junctions of the chloroplast reticulum and are tracked using a nearest-velocity tracker in trackpy \cite{trackpy} with a $(6\,\mathrm{\mu m},6\,\mathrm{\mu m},2\,\mathrm{\mu m})$  search window and $4$-time step memory ($\approx 40-60\,\mathrm{s}$), as well as an adaptive search window to down to $10\,\%$ of the initial search window size, to enhance the trajectory matching in a dense environment. 
Trajectories with less than $9$ timesteps, corresponding to about $90-120\,\mathrm{s}$, are rejected. We calculate velocities using a Savitzky-Golay Filter using a $30-45\,\mathrm{s}$ window to fit 2nd-order polynomials and take a smooth derivative.
The mean-squared displacement (MSD) of individual trajectories is calculated by a time average $\mathrm{MSD}(\tau)=\langle \vec{x}(t+\tau)\vec{x}(t) \rangle_t$ up to a maximal displacement of half the trajectories length ($\tau\leq T/2$).
Features such as Betti numbers and edge lengths are calculated on the graph.
The Euler characteristics are calculated from the mask using  region properties of scikit-image~\cite{van2014scikit}.



\subsection*{\large Supplementary Text I: Ecology and Light fluctuations}
Here we briefly discuss the typical light intensities and fluctuations experienced in the natural habitat of \textit{Pyrocystis lunula}.
Surface waves typically induce light fluctuations at frequencies of $0.1-1\,\mathrm{Hz}$~\cite{Stramska1998}, while clouds can cause more pronounced changes in light intensity over the course of minutes \cite{Darecki2011}.
\textit{Pyrocystis spp.} are known to live in the lower eutrophic zone at depths of $60-100\,\mathrm{m}$~\cite{Bhovichitra1977} and undergo strong diurnal vertical migration~\cite{Larson2022}. 
The growth of \textit{P. lunula} reaches full saturation under low light conditions found at approximately $50\,\mathrm{m}$ depth, equivalent to about $12\,\mathrm{mW/cm^2}$~\cite{Swift1976}.
These natural conditions therefore correspond to the regions in which \textit{P. lunula} adapts its chloroplast extension in response to varying light intensities (Fig.~\ref{fig:Figure1}D).



\subsection*{\large Supplementary Text II: Active Kelvin-Voigt model}
To mathematically describe light-induced chloroplast contraction and expansion, we employ a one-dimensional visco-elastic model.
This choice is grounded on the uni-axial contraction of the chloroplast network, with negligible expansion or contraction occurring in the perpendicular directions (Fig.~\ref{fig:Figure2}A,B,D, Movie S4-S6). We assume, that stresses are not externally imposed but applied along the chloroplast network itself. This assumption is supported by the observed structural similarity between actin-network and chloroplast structures~\cite{Heimann2009}. 
However, assuming constant rheological properties along the network and a linear force transmission, the picture of global stress application or local stresses is equivalent.
Viscous dissipation dampens the motion. Here, we assume this dissipation results either from a rheological feature (visco-elasticity) of the contracting network itself, the motion of the chloroplast network through the surrounding viscous cytoplasm, or from a combination of both mechanisms. The precise origin of this feature remains unspecified in our model.
Experimental measurements indicate that the fluctuations in chloroplast position are small compared to the global motion (Fig.~\ref{fig:Figure1}B,D, Movie S1-S3). This allows us to neglect noise in this model.
Thus, the stress-induced motion can be formalized as a visco-elastic Kelvin-Voigt solid:
\begin{align}
\label{eq:KV}
    \tau_{KV}\frac{\mathrm{d}x}{\mathrm{d}t} + x= f(t).
\end{align}
Here $x$ is a displacement from an equilibrium position $x_0=0$ and $f=F/k$ is an equilibrium position of the spring with spring constant $k$ forced with $F(t)$. $\tau_{KV}=\frac{\eta}{k}$ is a visco-elastic relaxation-timescale, where $\eta$ is a viscous damping factor.
If light $I(t)$ is sensed, the contractile force is gradually increased by a chemical concentration $c$: $f(t) = \beta c(t)$ hence the equilibrium position changes over time, until it reaches an intensity-dependent set-point. 
We model such a first-order reaction by a process with a timescale $\tau_1$  (fig.~\ref{fig:FigureA2}):
\begin{align}
\label{eq:phot2}
    \tau_1\frac{\mathrm{d}c_1(t)}{\mathrm{d}t} + c_1(t) = \alpha \, I(t)
\end{align}
with $\alpha$ such that $\mathrm{max}(c)=1$.
Moreover, we allow the relaxation timescale to take two different values $\tau_{KV}$ and $\tau_{KV}^*$, depending on whether light $I$ is ``on'' ($I>0$) or ``off'' ($I=0$)\footnote{Here ``on'' corresponds to blue or white light stimulation and ``off'' to dim red light in the experiments.}. Therefore, contraction and expansion will follow different timescales, which is likely a direct consequence of different rates of the underlying molecular contractile/extensile mechanism.
As we measure $x$ from the spread-out position $\mathcal{A}/\mathcal{A}_0=1$ we define: $\mathcal{A}(t)/\mathcal{A}_0 = 1-L*x(t)$ and incorporate $L$ in $\beta$ for convenience, without loss of generality.
These equations can be solved analytically for constant light irradiation starting at $t=0$ (as in our experiments Fig.~\ref{fig:Figure1}):
\begin{align}
x(t) = I\, \alpha\,\beta  \left[1-e^{-t/\tau_{KV}} + \frac{\tau_1}{\tau_{KV}-\tau_1}\left(e^{-t/\tau_1}-e^{-t/\tau_{KV}} \right)\right]
\end{align}

For simplicity we denote $\Delta \mathcal{A}_{max} \equiv \alpha\beta I$.

\subsubsection*{Transient response at low light $I<I_{th}$}
At low light intensities, we observe an adaptive response (Fig.~\ref{fig:Figure1}D). 
Similarly to models of light adaptation in green algae~\cite{Drescher2010}, we postulate that below a threshold intensity when $0< I< I_{th}$, a second light sensor sends an opposing signal (fig.~\ref{fig:FigureA2}), which controls the chloroplast spreading (accumulation response). The second sensor is modeled as
\begin{align}
\label{eq:phot1}
    \tau_2\frac{\mathrm{d}c_2(t)}{\mathrm{d}t} + c_2(t) = \alpha \, I(t).
\end{align}
Here the driving force is defined by the difference: $f=\beta(c_1-c_2)$. For times greater than $T_{max}=argmax(x)$ the direction of chloroplast motion reverses (expansion) and thus, as the time-scale of this outward-motion was found to be different,
equation \eqref{eq:transient} will then be used with $\tau_{KV}^*$.
The analytical solution for the case of $\tau_{KV} = \tau_{KV}^*$ reads:
\begin{align}
\label{eq:transient}
    x(t) = I\alpha \beta\left[\frac{\tau_2}{\tau_2-\tau_{KV}} \left( e^{-t/\tau_2}-e^{-t/\tau_{KV}}\right) +\frac{\tau_1}{\tau_{KV}-\tau_1} \left( e^{-t/\tau_1}-e^{-t/\tau_{KV}}\right) \right]
\end{align}

\subsubsection*{Dynamic filter-properties of chloroplast response}
The main driving factor of the observed behavior is the optimization of photosynthesis, while minimizing photodamage. We interpret such a response as a result of the cells signal-processing mechanism upon light sensing. Assuming there is a cost to such a response, the relevant light fluctuations have to be filtered out from the irrelevant ones. Consequently, our model can be viewed from the perspective of linear filters (if $I>I_{th}$).
We can easily see that the super-threshold dynamic equations \eqref{eq:KV} and \eqref{eq:phot1} will restore a harmonic oscillator by inserting equation \eqref{eq:KV} into \eqref{eq:phot2} with $f=\tau_{KV} c_1$ (simplifying $\tau_{KV}=\tau_{KV}^*$):
\begin{align}
\frac{\mathrm{d}^2x}{\mathrm{d}t^2} + \frac{\tau_1+\tau_{KV}}{\tau_1\tau_{KV}}\frac{\mathrm{d}x}{\mathrm{d}t} + \frac{x}{\tau_1\tau_{KV}} = \alpha I(t)
\end{align}
For any combination of the $\tau_{KV}$ and $\tau_1$ this system is over-damped as $0 < \frac{(\tau_1+\tau_{KV})^2}{4(\tau_1\tau_{KV})^2}-\frac{1}{\tau_1\tau_{KV}}$ \footnote{$0 < \frac{(\tau_1+\tau_{KV})^2}{4(\tau_1\tau_{KV})^2} -\frac{1}{\tau_1\tau_{KV}}
\;\Leftrightarrow\;0<\tau_1^2+2\tau_1\tau_{KV}+\tau_{KV}^2-4\tau_1\tau_{KV}=(\tau_1-\tau_{KV})^2 \;\square$}.
If our light-input is Fourier-transformable, i.e. $\hat I(\omega) = \int_{-\infty}^{\infty} I(t)e^{-i\omega t} \mathrm{d}t $ exists, we can find a linear response relationship $\hat{x}(\omega)= \hat{\xi}(\omega)\hat{I}(\omega)$, with a susceptibility $\xi(\omega) = \frac{\alpha}{(1-(\tau_1+\tau_{KV})i\omega)(\tau_1\tau_{KV})^{-1} -\omega^2}$, showing that the system essentially behaves as a long-pass filter, cutting off the high frequencies from noisy environmental fluctuations and adapting towards slow trends of light at time scales $\tau \approx 3-5\,\mathrm{min}$.

\subsection*{\large Model fitting}
We fit the models to every experiment and report the mean fit parameter and standard deviations. The $R^2$ score consistently exceeds $0.95$, indicating a reasonably good fit, despite the low complexity of the model and the noisy data.
The fitting scheme used is a least-squares fit.
The fitting parameters remained notably consistent for all different experiments and physiological conditions::
\begin{table}[hb]
\begin{tabular}{l|c|c|c|c|c|c}
   \textbf{Figure}  & \textbf{Irradiance}$\,\mathrm{[mW/cm^2]}$ & $\tau_1\,\mathrm{[min]}$ & $\tau_2\,\mathrm{[min]}$ & $\tau_{KV}\,\mathrm{[min]}$ &$\tau_{KV}^*\,\mathrm{[min]}$ & $\Delta \mathcal{A}_{max}$  \\\hline
    Fig.~1B, S4A-D & $2.8$& 1.1& -  &3.3&9&0.33\\
    Fig.~1D & $0.6$& $1.6\pm 0.5$ &$2.5\pm0.9$ & $2.9\pm 0.7$  & $6.2 \pm 2.5$& $0.40\pm0.12$\\
    Fig.~1D & $2.8$& $1.5\pm0.7$&- & $4.4\pm 0.9$&-& $0.23\pm0.04$\\
    Fig.~1D & $4.9$& $1.8\pm0.4$&-  &$3.6\pm1.3$&-&$0.43\pm 0.03$\\
    Fig.~1D & $7.6$& $1.7\pm0.5$&-  &$3.7\pm1.1$&-&$0.43\pm 0.06$\\
    Fig.~1D & $14.5$& $2.0\pm0.3$&-&$2.2\pm1.0$&-&$0.38\pm0.06$ \\
    Fig.~1D & $23.4$& $2.1\pm0.2$ &- &$2.4\pm0.2$&-&$0.36\pm0.05$ \\
    Fig.~1D & $32.4$&$1.9\pm0.7$ &- &$2.9\pm0.6$&- &$0.43\pm0.02$\\
    Fig.~1D &$41.4$&$2.0\pm0.2$ &- &$2.3\pm0.2$&-&$0.34\pm0.05$\\
    Fig.~3G & $3$    & $1.47\pm0.02$ & -  &  $1.47\pm 0.02$ &-& $0.65\pm0.01$ \\
    Fig.~4C top & $3.25$ & $2.5\pm0.5$ & $3.9\pm0.5$ & $2.8\pm 0.5$ &- & $1$ \\
    Fig.~4C bottom & $3.25$ &$2.1\pm 0.5$ &  - & $2.1\pm 0.5$ &-& $0.53\pm0.01$\\
    Fig.~4D top & $3.25$ &$1.9\pm 0.5$& - &  $1.9\pm 0.5$&-& $0.80\pm0.01$ \\
    Fig.~4D bottom & $3.25$ & $2.2\pm0.5$& - &$2.2\pm0.5$&-& $0.73\pm0.01$ \\
\end{tabular}
\end{table}

\pagebreak
\subsection*{\large Supplementary Text III: Pharmacological treatment}
Cells were treated with $5\,\mathrm{\mu M}$ Nocodazole (microtubule depolimerization), $10\,\mathrm{\mu M}$ Latrunculin B (actin depolimeization) or $2\,\mathrm{mM}$ BDM (myosin inhibition), all prepared in aqueous f/2 solution. Prior to treatment, cells were adapted to either darkness or bright light conditions ($I\approx 30\,\mathrm{mW/cm^2}$). Following the treatments, the cells were incubated for another $2\,\mathrm{h}$ in darkness, or bright light conditions ($I\approx 30\,\mathrm{mW/cm^2}$).
Microscopy is performed simultaneously for all treatment groups: first dark- and light-adapted cells are subjected to darkness for $70\,\mathrm{min}$. This was followed by exposure to intermediate light levels, with a subsequent increase in light intensity after $30\,\mathrm{min}$. \\
To measure the effect, we count the cells and measure the overall increase or decrease of the average pixel values with respect to the first time point. The data is normalized for the number of cells in the field of view, to account for size effects.\\
The unperturbed control-group of light adapted cells ($N=1588$) expands at dim light conditions within $50\,\mathrm{min}$ (fig.~\ref{fig:Figure_Pharma}A,B). BDM ($N=1138$) and Nocodazole ($N=1239$) treated cells do not change and expand as efficiently as the control group. Latrunculin B-treated cells ($N=1724$) do not expand their chloroplasts, as confirmed by the vanishing slope of the relative normalized absorbance (fig.~\ref{fig:Figure_Pharma}A) and by visual inspection (fig.~\ref{fig:Figure_Pharma}B). \\
For the group of dark adapted cells, which are subjected to intermediate light intensities ($\mathrm{Time}\leq30\,\mathrm{min}$) and strong light intensities ($\mathrm{Time}>30\,\mathrm{min}$), we find that all groups but the Latrunculin-B treated cells ($N=1724$) adapt to the strong light stimulus, as confirmed by the decline in absorbance and visual inspection (fig.~\ref{fig:Figure_Pharma}C,D). Our pharmacological inhibitions clearly suggests that the actin network is used both in chloroplast contraction and expansion. While the role of myosin and microtubules is not fully clear and might be concentration dependent, as suggested in studies on the diurnal chloroplast motion \cite{Heimann2009}.

\section*{SI Movies}
\textbf{Movie S1:} Chloroplast contraction of \textit{P. lunula} at different light intensities: (left) $0.6\,\mathrm{mW/cm^2}$, (center) $2.8\,\mathrm{mW/cm^2}$ and (right) $41.4\,\mathrm{mW/cm^2}$, corresponding to \ref{fig:FigureA1}A-C, respectively. \\
\textbf{Movie S2:} Dynamic light-controlled chloroplast motion with $10\,\mathrm{min}$ $0.4\,\mathrm{mW/cm^2}$ red-light imaging between $2.5\,\mathrm{min}$-lasting white-light stimulation at $I=7.6\,\mathrm{mW/cm^2}$.\\
\textbf{Movie S3:} Dynamic light-controlled chloroplast motion with $15\,\mathrm{min}$ $0.4\,\mathrm{mW/cm^2}$ red-light imaging between $15\,\mathrm{min}$-lasting white-light stimulation at $I=7.6\,\mathrm{mW/cm^2}$.\\
\textbf{Movie S4:} Time series for global stimulation with blue light ($470\pm 50\,\mathrm{nm}$. Imaging of chlorophyll auto-fluorescence (red look-up table (LUT)).\\
\textbf{Movie S5:} Time series for a second global stimulation with blue light ($470\pm 50\,\mathrm{nm}$. Imaging of chlorophyll auto-fluorescence (red look-up table (LUT)). Sequential buckling of cytoplasmic strands clearly visible.\\
\textbf{Movie S6:} Time series for a global stimulation with blue light ($470\pm 50\,\mathrm{nm}$. Imaging of chlorophyll auto-fluorescence (red look-up table (LUT)). After $14\,\mathrm{min}$ blue light is switched off and ambient red light is placed. Chloroplasts spread out within a larger time scale. Note the adjusted time step.\\
\textbf{Movie S7:} Network analysis of chloroplast autofluorescence signal. Nodes (dots) between the edges of the skeletonized image are tracked over time. Dataset corresponds to Fig. 3 and Movie S4.\\
\textbf{Movie S8:} Peripheral stimulation $488\,\mathrm{nm}$-laser (white box). Chloroplast auto fluorescence (red LUT).\\
\textbf{Movie S9:} Sequential peripheral stimulation $488\,\mathrm{nm}$-laser (white box). Chloroplast auto fluorescence (red LUT). Chloroplast shrinks first on upper stimulation side, while simultaneously the lower side reacts. Then a second stimulus is applied to the lower side.\\
\textbf{Movie S10:} Central stimulation with $488\,\mathrm{nm}$-laser (white box). Chloroplast auto fluorescence (red LUT).\\
\newpage
\section*{SI Figures}
\setcounter{figure}{0}
\renewcommand{\figurename}{Fig.}
\renewcommand{\thefigure}{S\arabic{figure}}

\begin{figure}[h!]
    \centering
    \includegraphics[width=1\textwidth]{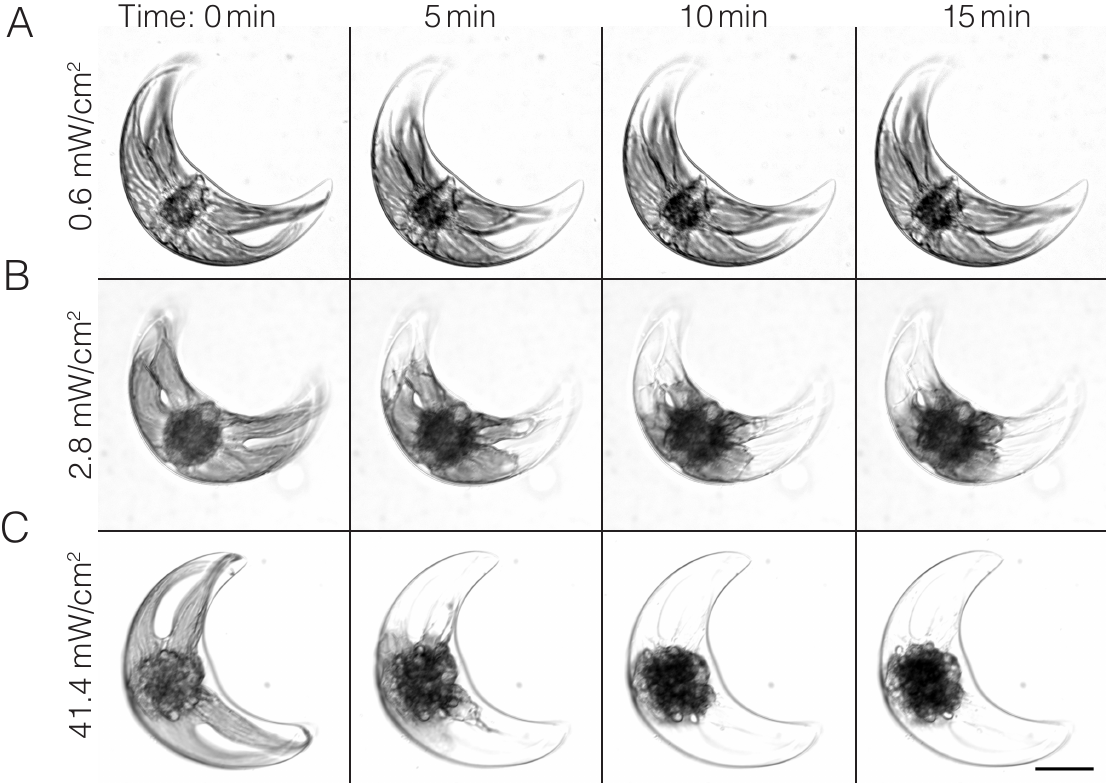}
    \caption{\textbf{Irradiance-dependent chloroplast retraction.} (\textbf{A}) at low intensities, chloroplast contraction is limited to a transient response. (\textbf{B}) Intermediate intensities lead to an incomplete retraction of the chloroplast. 
(\textbf{C}) High-intensity light stimulation leads to a complete retraction of the chloroplast towards the cytoplasmic core area. Scale bar: $30\,\mathrm{\mu m}$}
    \label{fig:FigureA1}
\end{figure}

\begin{figure}[h!]
    \centering
    \includegraphics[width=.6\textwidth]{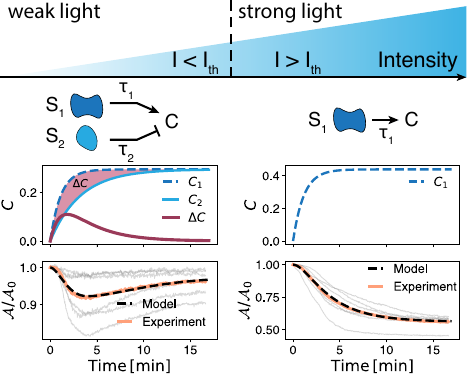}
    \caption{\textbf{Schematic representation of irradiance-dependent signaling model}. Left: for small intensities, light reacts with two opposing light sensors ($S_1$ and $S_2$) with two different time scales $\tau_1<\tau_2$, leading to a non-monotonic curve of the concentration $\Delta C$ (magenta line), and hence to a non-monotonic change in area of the chloroplast (dotted black line, lower graph).
    As the light intensity is larger than a threshold $I\geq I_{th}$ the second sensor $S_2$ is not opposing the signal of sensor $S_1$ any further. Hence the response is purely photoavoidant (monotonous decrease of chloroplast area $\mathcal{A}$).}
    \label{fig:FigureA2}
\end{figure}

\begin{figure}[h!]
    \centering
    \includegraphics[width=.6\textwidth]{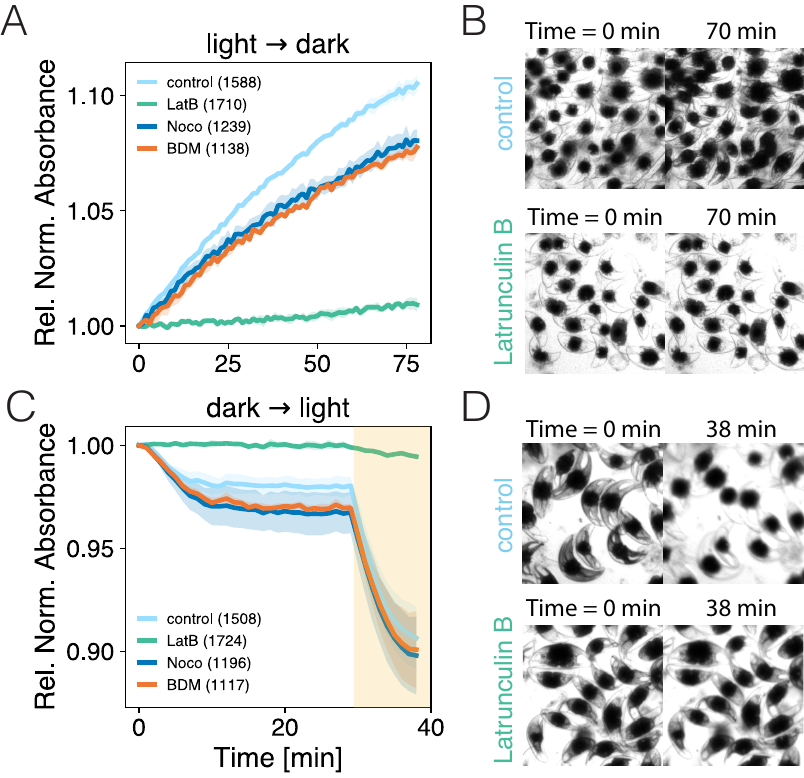}
    \caption{Analysis of pharmacological perturbation to the photoadaptation mechanism. (\textbf{A}) Light-adapted cells are placed in dark light conditions and spread out (control). Absorption is measured and normalized with the amount of cells per experiment. $10\,\mathrm{\mu M}$ Latrunculin B treated cells (LatB) do not react. Both $5\,\mathrm{\mu M}$ Nocodazole (Noco) and $2\,\mathrm{mM}$ 2,3-butanedione monoxime (BDM) do not show a significant effect. Numbers in the legend stand for the total amount of cells per treatment. Shadowed regions display the min-max error of two independent batches. (\textbf{B}) Example images for control and Latrunculin B treated cells before and after dark adaptation. 
    (\textbf{C}) Absorption measurement of dim light-adapted cells placed in bright light, with subsequent increase of light intensity (shaded area).
    (\textbf{D}) Example for control and Latrunculin B treated cells before and after bright light adaptation. }
    \label{fig:Figure_Pharma}
\end{figure}

\begin{figure}[h!]
    \centering
    \includegraphics[width=1\textwidth]{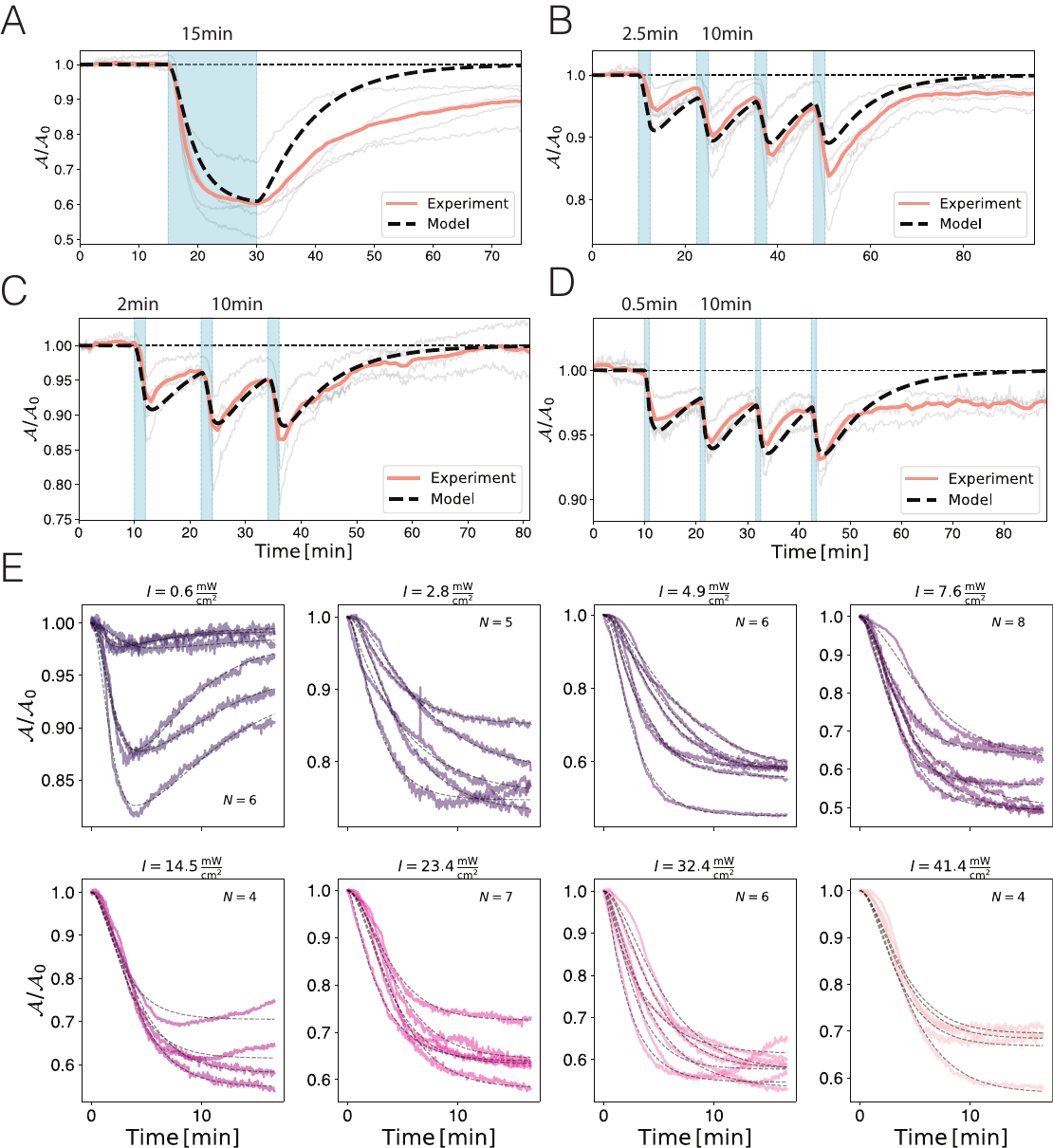}
    \caption{(A-D) Integration of dynamical model (for $I>I_{th}$) (dotted line) compared to experimental data (orange: mean) for different durations of light illumination (blue region) and dim red light. Parameters for the model were $\tau_1 = 1.1\,\mathrm{min}$, $\tau_{KV}=3.3\,\mathrm{min}$, $\tau_{KV}^* = 9\,\mathrm{min}$ (Table 1). At long times the area was underestimated. (E) Fits of all individual experiments at different light intensities (colors).}
    \label{fig:SI_Figure_AllFits}
\end{figure}

\clearpage
\newpage

\printbibliography

@article{Swift1967,
author = {Swift, Elijah and Taylor, W. Rowland},
doi = {10.1111/j.1529-8817.1967.tb04634.x},
issn = {15298817},
journal = {Journal of Phycology},
number = {2},
pages = {77--81},
title = {{Bioluminescence and chloroplast movement in the dinoflagellate Pyrocystis lunula}},
volume = {3},
year = {1967}
}

@book{Senn1908,
author = {Senn, Gustav},
booktitle = {Verlag von Wilhelm Engelmann},
title = {{Die Gestalts- und Lagever{\"{a}}nderung der Pflanzen-Chromatophoren.}},
publisher = {Wilhelm Engelmann, Leipzig},
year = {1908}
}

@article{Bhaya2004,
author = {Bhaya, Devaki},
doi = {10.1111/j.1365-2958.2004.04160.x},
issn = {0950382X},
journal = {Molecular Microbiology},
number = {3},
pages = {745--754},
pmid = {15255889},
title = {{Light matters: Phototaxis and signal transduction in unicellular cyanobacteria}},
volume = {53},
year = {2004}
}

@article{Sagan1967,
author = {Sagan, Lynn},
doi = {10.1016/0022-5193(67)90079-3},
issn = {10958541},
journal = {Journal of Theoretical Biology},
number = {3},
pages = {225--274},
pmid = {11541392},
title = {{On the origin of mitosing cells}},
volume = {14},
year = {1967}
}

@article{Biggley1969,
author = {Biggley, W.H. and Swift, E. and Buchanan, R. J. and Seliger, H.H.},
journal = {The Journal of general physiology},
pages = {96--122},
title = {{Stimulable and Spontaneous Bioluminescence in the Marine Dinoflagellates , Pyrodinium and Pyrocystis lunula}},
volume = {54},
year = {1969},
doi = {10.1085/jgp.54.1.96}
}

@article{fortes1989poison,
title = {The poison effect in cork},
journal = {Materials Science and Engineering: A},
volume = {122},
number = {2},
pages = {227-232},
year = {1989},
issn = {0921-5093},
doi = {https://doi.org/10.1016/0921-5093(89)90634-5},
url = {https://www.sciencedirect.com/science/article/pii/0921509389906345},
author = {M.A. Fortes and M. {Teresa Nogueira}},

}

@article{Ruban2015,
author = {Ruban, Alexander V.},
doi = {10.1093/jxb/eru400},
issn = {14602431},
journal = {Journal of Experimental Botany},
number = {1},
pages = {7--23},
pmid = {25336689},
title = {{Evolution under the sun: Optimizing light harvesting in photosynthesis}},
volume = {66},
year = {2015}
}

@article{Menghini2021,
doi = {10.1016/j.dib.2021.107254},
issn = {23523409},
journal = {Data in Brief},
pages = {107254},
publisher = {Elsevier Inc.},
title = {{De novo transcriptome assembly data of the marine bioluminescent dinoflagellate Pyrocystis lunula}},
url = {https://doi.org/10.1016/j.dib.2021.107254},
volume = {37},
year = {2021}
}

@article{Moulton2020,
author = {Moulton, Derek E. and Oliveri, Hadrien and Goriely, Alain},
doi = {10.1073/pnas.2016025117},
issn = {10916490},
journal = {Proceedings of the National Academy of Sciences of the United States of America},
number = {51},
pages = {32226--32237},
pmid = {33273121},
title = {{Multiscale integration of environmental stimuli in plant tropism produces complex behaviors}},
volume = {117},
year = {2020}
}

@article{Fast2001,
author = {Fast, Naomi M. and Kissinger, Jessica C. and Roos, David S. and Keeling, Patrick J.},
doi = {10.1093/oxfordjournals.molbev.a003818},
issn = {07374038},
journal = {Molecular Biology and Evolution},
number = {3},
pages = {418--426},
pmid = {11230543},
title = {{Nuclear-encoded, plastid-targeted genes suggest a single common origin for apicomplexan and dinoflagellate plastids}},
volume = {18},
year = {2001}
}

@article{Gray1992,
author = {Gray, Michael W.},
doi = {10.1016/S0074-7696(08)62068-9},
issn = {00747696},
journal = {International Review of Cytology},
number = {C},
pages = {233--357},
pmid = {1452433},
title = {{The Endosymbiont Hypothesis Revisited}},
volume = {141},
year = {1992}
}

@article{Bhattacharya2004,
author = {Bhattacharya, Debashish and Yoon, Hwan Su and Hackett, Jeremiah D.},
doi = {10.1002/bies.10376},
issn = {02659247},
journal = {BioEssays},
number = {1},
pages = {50--60},
pmid = {14696040},
title = {{Photosynthetic eukaryotes unite: Endosymbiosis connects the dots}},
volume = {26},
year = {2004}
}

@article{Riviere2023,
author = {Rivi{\`{e}}re, Mathieu and Meroz, Yasmine},
doi = {10.1073/pnas.2306655120},
issn = {10916490},
journal = {Proceedings of the National Academy of Sciences of the United States of America},
number = {42},
pmid = {37816057},
title = {{Plants sum and subtract stimuli over different timescales}},
volume = {120},
year = {2023}
}

@article{Meroz2019,
author = {Meroz, Yasmine and Bastien, Renaud and Mahadevan, L.},
doi = {10.1098/rsif.2019.0038},
issn = {17425662},
journal = {Journal of the Royal Society Interface},
number = {154},
pmid = {31088258},
title = {{Spatio-temporal integration in plant tropisms}},
volume = {16},
year = {2019}
}

@article{Larson2022,
	author = {Adam G. Larson and Rahul Chajwa and Hongquan Li and Manu Prakash},
	title = {Inflation induced motility for long-distance vertical migration},
	elocation-id = {2022.08.19.504465},
	year = {2023},
	doi = {10.1101/2022.08.19.504465},
	publisher = {Cold Spring Harbor Laboratory},
	URL = {https://www.biorxiv.org/content/early/2023/11/14/2022.08.19.504465},
	eprint = {https://www.biorxiv.org/content/early/2023/11/14/2022.08.19.504465.full.pdf},
	journal = {bioRxiv}
}

@article{Park1996,
author = {Park, Youn Il and Chow, Wah Soon and Andersen, Jan M.},
doi = {10.1104/pp.111.3.867},
isbn = {4690166676},
issn = {00320889},
journal = {Plant Physiology},
number = {3},
pages = {867--875},
pmid = {12226333},
title = {{Chloroplast Movement in the Shade Plant Tradescantia albiflora Helps Protect Photosystem II against Light Stress}},
volume = {111},
year = {1996}
}

@article{Bhovichitra1977,
author = {Bhovichitra, Mahn and Swift, Elijah},
doi = {10.4319/lo.1977.22.1.0073},
issn = {19395590},
journal = {Limnology and Oceanography},
number = {1},
pages = {73--83},
title = {{Light and dark uptake of nitrate and ammonium by large oceanic dinoflagellates: Pyrocystis noctiluca, Pyrocystis fusiformis, and Dissodinium lunula}},
volume = {22},
year = {1977}
}

@article{Li2009,
author = {Li, Zhirong and Wakao, Setsuko and Fischer, Beat B. and Niyogi, Krishna K.},
doi = {10.1146/annurev.arplant.58.032806.103844},
issn = {15435008},
journal = {Annual Review of Plant Biology},
pages = {239--260},
pmid = {19575582},
title = {{Sensing and responding to excess light}},
volume = {60},
year = {2009}
}

@article{Kasahara2002,
author = {Kasahara, Masahiro and Kagawa, Takatoshi and Olkawa, Kazusato and Suetsugu, Noriyuki and Miyao, Mitsue and Wada, Masamitsu},
doi = {10.1038/nature01213},
issn = {00280836},
journal = {Nature},
number = {6917},
pages = {829--832},
pmid = {12490952},
title = {{Chloroplast avoidance movement reduces photodamage in plants}},
volume = {420},
year = {2002}
}

@article{Haupt1982,
author = {Haupt, Wolfgang},
journal = {Ann. Rev. Plant Physiol.},
pages = {205--233},
title = {{Light-mediated movement of chloroplasts}},
url = {https://doi.org/10.1146/annurev.pp.33.060182.001225},
volume = {33},
year = {1982}
}

@article{Darecki2011,
author = {Darecki, Miroslaw and Stramski, Dariusz and Sok{\'{o}}lski, MacIej},
doi = {10.1029/2011JC007338},
issn = {21699291},
journal = {Journal of Geophysical Research: Oceans},
number = {11},
pages = {1--16},
title = {{Measurements of high-frequency light fluctuations induced by sea surface waves with an Underwater Porcupine Radiometer System}},
volume = {116},
year = {2011}
}

@article{Stramska1998,
author = {Stramska, Malgorzata and Dickey, Tommy D.},
doi = {10.1016/S0967-0637(98)00020-X},
issn = {09670637},
journal = {Deep-Sea Research Part I: Oceanographic Research Papers},
number = {9},
pages = {1393--1410},
title = {{Short-term variability of the underwater light field in the oligotrophic ocean in response to surface waves and clouds}},
volume = {45},
year = {1998}
}

@article{Swift1976,
author = {Swift, Elijah and Meunier, Valerie},
year = {1976},
pages = {14 - 22},
title = {Effects of light intensity on division rate, stimulable bioluminescence and cell size of the oceanic dinoflagellates Dissodinium lunula, Pyrocystis fusiformis and P. noctiluca},
volume = {12},
number = {1},
journal = {Journal of Phycology},
doi = {10.1111/j.1529-8817.1976.tb02819.x}
}

@article{Jekely2009,
author = {J{\'{e}}kely, G{\'{a}}sp{\'{a}}r},
doi = {10.1098/rstb.2009.0072},
issn = {14712970},
journal = {Philosophical Transactions of the Royal Society B: Biological Sciences},
number = {1531},
pages = {2795--2808},
pmid = {19720645},
title = {{Evolution of phototaxis}},
volume = {364},
year = {2009}
}

@article{Topperwien1980,
author = {T{\"{o}}pperwien, F. and Hardeland, R.},
doi = {10.1080/09291018009359717},
issn = {0022-1945},
journal = {Journal of Interdisciplinary Cycle Research},
number = {4},
pages = {325--329},
title = {{Free‐running circadian rhythm of plastid movements in individual cells of pyrocystis lunula (dinophyta) }},
volume = {11},
year = {1980}
}

@article{edelstein2014,
  title={Advanced methods of microscope control using $\mu$Manager software},
  author={Edelstein, Arthur D and Tsuchida, Mark A and Amodaj, Nenad and Pinkard, Henry and Vale, Ronald D and Stuurman, Nico},
  journal={Journal of biological methods},
  volume={1},
  number={2},
  year={2014},
  publisher={NIH Public Access}
}

@article{Schuergers2016,
article_type = {journal},
title = {Cyanobacteria use micro-optics to sense light direction},
author = {Schuergers, Nils and Lenn, Tchern and Kampmann, Ronald and Meissner, Markus V and Esteves, Tiago and Temerinac-Ott, Maja and Korvink, Jan G and Lowe, Alan R and Mullineaux, Conrad W and Wilde, Annegret},
editor = {Rieke, Fred},
volume = {5},
year = {2016},
pages = {e12620},
citation = {eLife 2016;5:e12620},
doi = {10.7554/eLife.12620},
url = {https://doi.org/10.7554/eLife.12620},
journal = {eLife},
issn = {2050-084X},
publisher = {eLife Sciences Publications, Ltd},
}

@article{Wada2013,
author = {Wada, Masamitsu},
doi = {10.1016/j.plantsci.2013.05.016},
issn = {01689452},
journal = {Plant Science},
pages = {177--182},
pmid = {23849124},
publisher = {Elsevier Ireland Ltd},
title = {{Chloroplast movement}},
url = {http://dx.doi.org/10.1016/j.plantsci.2013.05.016},
volume = {210},
year = {2013}
}

@article{Swift1970,
author = {Swift, Elijah and Remsen, Charles C.},
title = {The cell wall of Pyrocystis SPP (Dinococcales)},
journal = {Journal of Phycology},
volume = {6},
number = {1},
pages = {79-86},
doi = {https://doi.org/10.1111/j.1529-8817.1970.tb02361.x},
url = {https://onlinelibrary.wiley.com/doi/abs/10.1111/j.1529-8817.1970.tb02361.x},
eprint = {https://onlinelibrary.wiley.com/doi/pdf/10.1111/j.1529-8817.1970.tb02361.x},
year = {1970}
}

@article{Heimann2009,
author = {Heimann, Kirsten and Klerks, Paul L. and Hasenstein, Karl H.},
doi = {10.1515/BOT.2009.010},
issn = {00068055},
journal = {Botanica Marina},
number = {2},
pages = {170--177},
title = {{Involvement of actin and microtubules in regulation of bioluminescence and translocation of chloroplasts in the dinoflagellate Pyrocystis lunula}},
volume = {52},
year = {2009}
}

@article{Krause1991,
author = {Krause, G.H. and Weis, E.},
doi = {10.1002/9780470122563.ch3},
isbn = {9780470122563},
journal = {Annual Review of Plant Physiology and Plant Molecular Biology},
pages = {313--349},
title = {{Chlorophyll fluorescence and photosynthesis}},
volume = {42},
year = {1991}
}

@article{Surpin2002,
author = {Surpin, Marci and Larkin, Robert M. and Chory, Joanne},
doi = {10.1105/tpc.010446},
issn = {10404651},
journal = {Plant Cell},
number = {SUPPL.},
pages = {327--338},
pmid = {12045286},
title = {{Signal transduction between the chloroplast and the nucleus}},
volume = {14},
year = {2002}
}

@article{Jin2013,
author = {Jin, Kelly and Klima, Jason C and Deane, Grant and {Dale Stokes}, Malcolm and Latz, Michael I},
doi = {10.1111/jpy.12084},
issn = {00223646},
journal = {Journal of Phycology},
number = {4},
pages = {733--745},
title = {{Pharmacological investigation of the bioluminescence signaling pathway of the dinoflagellate Lingulodinium polyedrum: Evidence for the role of stretch-activated ion channels}},
volume = {49},
year = {2013}
}

@article{Harada2007,
author = {Harada, Akiko and Shimazaki, Ken-ichiro},
doi = {10.1562/2006-03-08-ir-837},
issn = {00318655},
journal = {Photochemistry and Photobiology},
number = {1},
pages = {102--111},
pmid = {16906793},
title = {{Phototropins and Blue Light-dependent Calcium Signaling in Higher Plants†}},
volume = {83},
year = {2007}
}

@book{Darwin1880,
  title={The power of movement in plants},
  author={Darwin, Charles and Darwin, Francis},
  year={1880},
  publisher={John Murray}
}

@article{Suetsugu2007,
author = {Suetsugu, Noriyuki and Wada, Masamitsu},
doi = {10.1515/BC.2007.118},
journal = {Biological Chemistry},
number = {9},
pages = {927--935},
title = {{Chloroplast photorelocation movement mediated by phototropin family proteins in green plants}},
volume = {388},
year = {2007}
}

@article{Liscum2014,
author = {Liscum, Emmanuel and Askinosie, Scott K. and Leuchtman, Daniel L. and Morrow, Johanna and Willenburg, Kyle T. and Coats, Diana Roberts},
doi = {10.1105/tpc.113.119727},
isbn = {0000000201490},
issn = {1532298X},
journal = {Plant Cell},
number = {1},
pages = {38--55},
pmid = {24481074},
title = {{Phototropism: Growing towards an understanding of plant movement}},
volume = {26},
year = {2014}
}

@article{Cohen2021,
doi = {10.1038/s41564-020-00814-7},
issn = {20585276},
journal = {Nature Microbiology},
number = {2},
pages = {173--186},
pmid = {33398100},
publisher = {Springer US},
title = {{Dinoflagellates alter their carbon and nutrient metabolic strategies across environmental gradients in the central Pacific Ocean}},
url = {http://dx.doi.org/10.1038/s41564-020-00814-7},
volume = {6},
year = {2021}
}

@article{Lin2011,
author = {Lin, Senjie},
doi = {10.1016/j.resmic.2011.04.006},
issn = {09232508},
journal = {Research in Microbiology},
number = {6},
pages = {551--569},
publisher = {Elsevier Masson SAS},
title = {{Genomic understanding of dinoflagellates}},
url = {http://dx.doi.org/10.1016/j.resmic.2011.04.006},
volume = {162},
year = {2011}
}

@article{Wada2018,
author = {Wada, Masamitsu and Kong, Sam Geun},
doi = {10.1242/jcs.210310},
issn = {14779137},
journal = {Journal of Cell Science},
number = {2},
pmid = {29378837},
title = {{Actin-mediated movement of chloroplasts}},
volume = {131},
year = {2018}
}

@article{Banas2012,
author = {Bana{\'{s}}, Agnieszka Katarzyna and Aggarwal, Chhavi and {\L}abuz, Justyna and Sztatelman, Olga and Gabry{\'{s}}, Halina},
doi = {10.1093/jxb/err429},
issn = {00220957},
journal = {Journal of Experimental Botany},
number = {4},
pages = {1559--1574},
pmid = {22312115},
title = {{Blue light signalling in chloroplast movements}},
volume = {63},
year = {2012}
}

@article{DeMaleprade2020,
archivePrefix = {arXiv},
arxivId = {1911.08837},
author = {{De Maleprade}, H{\'{e}}l{\`{e}}ne and Moisy, Fr{\'{e}}d{\'{e}}ric and Ishikawa, Takuji and Goldstein, Raymond E},
doi = {10.1103/PhysRevE.101.022416},
eprint = {1911.08837},
issn = {24700053},
journal = {Physical Review E},
number = {2},
pmid = {32168596},
title = {{Motility and phototaxis of Gonium, the simplest differentiated colonial alga}},
volume = {101},
year = {2020}
}

@article{Jin2020,

author = {Jin, Di and Kotar, Jurij and Silvester, Emma and Leptos, Kyriacos C. and Croze, Ottavio A.},
doi = {10.1016/j.bpj.2020.10.006},
issn = {15420086},
journal = {Biophysical Journal},
number = {10},
pages = {2055--2062},
publisher = {Biophysical Society},
title = {{Diurnal Variations in the Motility of Populations of Biflagellate Microalgae}},
url = {https://doi.org/10.1016/j.bpj.2020.10.006},
volume = {119},
year = {2020}
}

@article{Drescher2010,
author = {Drescher, Knut and Goldstein, Raymond E and Tuval, Idan},
doi = {10.1073/pnas.1000901107},
title = {{Fidelity of adaptive phototaxis}},
volume = {2010},
year = {2010}
}

@article{Tesson2015,
title = {Mechanosensitivity of a Rapid Bioluminescence Reporter System Assessed by Atomic Force Microscopy},
journal = {Biophysical Journal},
volume = {108},
number = {6},
pages = {1341-1351},
year = {2015},
issn = {0006-3495},
doi = {https://doi.org/10.1016/j.bpj.2015.02.009},
url = {https://www.sciencedirect.com/science/article/pii/S0006349515001691},
author = {Benoit Tesson and Michael I. Latz},
}

@article{Jalaal2020,
  title = {Stress-Induced Dinoflagellate Bioluminescence at the Single Cell Level},
  author = {Jalaal, Maziyar and Schramma, Nico and Dode, Antoine and de Maleprade, H\'el\`ene and Raufaste, Christophe and Goldstein, Raymond E.},
  journal = {Phys. Rev. Lett.},
  volume = {125},
  issue = {2},
  pages = {028102},
  numpages = {6},
  year = {2020},
  publisher = {American Physical Society},
  doi = {10.1103/PhysRevLett.125.028102},
  url = {https://link.aps.org/doi/10.1103/PhysRevLett.125.028102}
}

@article{Schramma2023,
author = {Schramma, Nico and Isra{\"{e}}ls, Cintia Perugachi and Jalaal, Maziyar},
doi = {10.1073/pnas.2216497120/-/DCSupplemental.Published},
eprint = {2204.07386},
journal = {Proceedings of the National Academy of Sciences},
number = {3},
pages = {1--9},
title = {{Chloroplasts in plant cells show active glassy behavior under low light conditions}},
url = {http://arxiv.org/abs/2204.07386},
volume = {120},
year = {2023}
}

@article{Pfreundt2023,
author = {Pfreundt, Ulrike and S{\l}omka, Jonasz and Schneider, Giulia and Sengupta, Anupam and Carrara, Francesco and Fernandez, Vicente and Ackermann, Martin and Stocker, Roman},
doi = {10.1126/science.adf2753},
issn = {10959203},
journal = {Science},
number = {6647},
pages = {830--835},
pmid = {37228200},
title = {{Controlled motility in the cyanobacterium Trichodesmium regulates aggregate architecture}},
volume = {380},
year = {2023}
}

@article{Fritz2000,
author = {Seo, Kyung Suk and Fritz, Lawrence},
doi = {10.1046/j.1529-8817.2000.99196.x},
issn = {00223646},
journal = {Journal of Phycology},
number = {2},
pages = {351--358},
title = {{Cell ultrastructural changes correlate with circadian rhythms in Pyrocystis lunula (Pyrrophyta)}},
volume = {36},
year = {2000}
}

@article{van2014scikit,
 title = {scikit-image: image processing in Python},
 author = {van der Walt, Stéfan and Schönberger, Johannes L. and Nunez-Iglesias, Juan and Boulogne, François and Warner, Joshua D. and Yager, Neil and Gouillart, Emmanuelle and Yu, Tony and the scikit-image contributors},
 year = 2014,
 month = jun,
 volume = 2,
 pages = {e453},
 journal = {PeerJ},
 issn = {2167-8359},
 url = {https://doi.org/10.7717/peerj.453},
 doi = {10.7717/peerj.453}
}

@article{NetworkX,
title = {Exploring network structure, dynamics, and function using NetworkX},
author = {Hagberg, Aric and Swart, Pieter J. and Schult, Daniel A.},
abstractNote = {NetworkX is a Python language package for exploration and analysis of networks and network algorithms. The core package provides data structures for representing many types of networks, or graphs, including simple graphs, directed graphs, and graphs with parallel edges and self loops. The nodes in NetworkX graphs can be any (hashable) Python object and edges can contain arbitrary data; this flexibility mades NetworkX ideal for representing networks found in many different scientific fields. In addition to the basic data structures many graph algorithms are implemented for calculating network properties and structure measures: shortest paths, betweenness centrality, clustering, and degree distribution and many more. NetworkX can read and write various graph formats for eash exchange with existing data, and provides generators for many classic graphs and popular graph models, such as the Erdoes-Renyi, Small World, and Barabasi-Albert models, are included. The ease-of-use and flexibility of the Python programming language together with connection to the SciPy tools make NetworkX a powerful tool for scientific computations. We discuss some of our recent work studying synchronization of coupled oscillators to demonstrate how NetworkX enables research in the field of computational networks.},
doi = {},
url = {https://www.osti.gov/biblio/960616}, journal = {},
number = {},
volume = {},
place = {United States},
year = {2008},
month = {1}
}

@article{Napari,
  author       = {Ahlers, Jannis and
                  Althviz Moré, Daniel and
                  Amsalem, Oren and
                  Anderson, Ashley and
                  Bokota, Grzegorz and
                  Boone, Peter and
                  Bragantini, Jordão and
                  Buckley, Genevieve and
                  Burt, Alister and
                  Bussonnier, Matthias and
                  Can Solak, Ahmet and
                  Caporal, Clément and
                  Doncila Pop, Draga and
                  Evans, Kira and
                  Freeman, Jeremy and
                  Gaifas, Lorenzo and
                  Gohlke, Christoph and
                  Gunalan, Kabilar and
                  Har-Gil, Hagai and
                  Harfouche, Mark and
                  Harrington, Kyle I. S. and
                  Hilsenstein, Volker and
                  Hutchings, Katherine and
                  Lambert, Talley and
                  Lauer, Jessy and
                  Lichtner, Gregor and
                  Liu, Ziyang and
                  Liu, Lucy and
                  Lowe, Alan and
                  Marconato, Luca and
                  Martin, Sean and
                  McGovern, Abigail and
                  Migas, Lukasz and
                  Miller, Nadalyn and
                  Muñoz, Hector and
                  Müller, Jan-Hendrik and
                  Nauroth-Kreß, Christopher and
                  Nunez-Iglesias, Juan and
                  Pape, Constantin and
                  Pevey, Kim and
                  Peña-Castellanos, Gonzalo and
                  Pierré, Andrea and
                  Rodríguez-Guerra, Jaime and
                  Ross, David and
                  Royer, Loic and
                  Russell, Craig T. and
                  Selzer, Gabriel and
                  Smith, Paul and
                  Sobolewski, Peter and
                  Sofiiuk, Konstantin and
                  Sofroniew, Nicholas and
                  Stansby, David and
                  Sweet, Andrew and
                  Vierdag, Wouter-Michiel and
                  Wadhwa, Pam and
                  Weber Mendonça, Melissa and
                  Windhager, Jonas and
                  Winston, Philip and
                  Yamauchi, Kevin},
  title        = {{napari: a multi-dimensional image viewer for 
                   Python}},
  month        = jul,
  year         = 2023,
  publisher    = {Zenodo},
  version      = {v0.4.18},
  doi          = {10.5281/zenodo.8115575},
  url          = {https://doi.org/10.5281/zenodo.8115575}
}

@article{Behrenfeld2014,
author = {Behrenfeld, Michael J.},
doi = {10.1038/nclimate2349},
issn = {17586798},
journal = {Nature Climate Change},
number = {10},
pages = {880--887},
title = {{Climate-mediated dance of the plankton}},
volume = {4},
year = {2014}
}

@article{Cavicchioli2019,
author = {Cavicchioli, Ricardo and Ripple, William J. and Timmis, Kenneth N. and Azam, Farooq and Bakken, Lars R. and Baylis, Matthew and Behrenfeld, Michael J. and Boetius, Antje and Boyd, Philip W. and Classen, Aim{\'{e}}e T. and Crowther, Thomas W. and Danovaro, Roberto and Foreman, Christine M. and Huisman, Jef and Hutchins, David A. and Jansson, Janet K. and Karl, David M. and Koskella, Britt and {Mark Welch}, David B. and Martiny, Jennifer B.H. and Moran, Mary Ann and Orphan, Victoria J. and Reay, David S. and Remais, Justin V. and Rich, Virginia I. and Singh, Brajesh K. and Stein, Lisa Y. and Stewart, Frank J. and Sullivan, Matthew B. and van Oppen, Madeleine J.H. and Weaver, Scott C. and Webb, Eric A. and Webster, Nicole S.},
doi = {10.1038/s41579-019-0222-5},
issn = {17401534},
journal = {Nature Reviews Microbiology},
number = {9},
pages = {569--586},
pmid = {31213707},
publisher = {Springer US},
title = {{Scientists' warning to humanity: microorganisms and climate change}},
url = {http://dx.doi.org/10.1038/s41579-019-0222-5},
volume = {17},
year = {2019}
}

@article{Field1998,
author = {Field, Christopher B. and Behrenfeld, Michael J. and Randerson, James T. and Falkowski, Paul},
doi = {10.1126/science.281.5374.237},
issn = {00368075},
journal = {Science},
number = {5374},
pages = {237--240},
pmid = {9657713},
title = {{Primary production of the biosphere: Integrating terrestrial and oceanic components}},
volume = {281},
year = {1998}
}

@article{Williamson1993,
author = {Williamson, Richard E},
doi = {10.1146/annurev.pp.44.060193.001145},
issn = {10402519},
journal = {Annual Review of Plant Physiology and Plant Molecular Biology},
number = {1},
pages = {181--202},
title = {{Organelle movements}},
volume = {44},
year = {1993}
}

@article{lee1994building,
  title={Building skeleton models via 3-D medial surface axis thinning algorithms},
  author={Lee, Ta-Chih and Kashyap, Rangasami L and Chu, Chong-Nam},
  journal={CVGIP: Graphical Models and Image Processing},
  volume={56},
  number={6},
  pages={462--478},
  year={1994},
  publisher={Elsevier}
}

@article{Gray2017,
author = {Gray, Michael W.},
doi = {10.1091/mbc.E16-07-0509},
issn = {19394586},
journal = {Molecular Biology of the Cell},
number = {10},
pages = {1285--1287},
pmid = {28495966},
title = {{Lynn Margulis and the endosymbiont hypothesis: 50 years later}},
volume = {28},
year = {2017}
}

@article{Florijn2014,
author = {Florijn, Bastiaan and Coulais, Corentin and {Van Hecke}, Martin},
doi = {10.1103/PhysRevLett.113.175503},
eprint = {1407.4273},
issn = {10797114},
journal = {Physical Review Letters},
number = {17},
pages = {1--5},
title = {{Programmable mechanical metamaterials}},
volume = {113},
year = {2014}
}

@article{Domaschke2019,
author = {Domaschke, S. and Morel, A. and Fortunato, G. and Ehret, A. E.},
doi = {10.1038/s41467-019-12757-7},
isbn = {4146701912757},
issn = {20411723},
journal = {Nature Communications},
number = {1},
pages = {1--8},
pmid = {31653833},
publisher = {Springer US},
title = {{Random auxetics from buckling fibre networks}},
url = {http://dx.doi.org/10.1038/s41467-019-12757-7},
volume = {10},
year = {2019}
}

@article{Bertoldi2017,
author = {Bertoldi, Katia and Vitelli, Vincenzo and Christensen, Johan and {Van Hecke}, Martin},
booktitle = {Nature Reviews Materials},
doi = {10.1038/natrevmats.2017.66},
issn = {20588437},
title = {{Flexible mechanical metamaterials}},
volume = {2},
year = {2017}
}

@article{Lakes1987,
author = {Roderic Lakes },
title = {Foam Structures with a Negative Poisson's Ratio},
journal = {Science},
volume = {235},
number = {4792},
pages = {1038-1040},
year = {1987},
doi = {10.1126/science.235.4792.1038},
URL = {https://www.science.org/doi/abs/10.1126/science.235.4792.1038},
eprint = {https://www.science.org/doi/pdf/10.1126/science.235.4792.1038}}

@article{Briggs2002,
author = {Briggs, Winslow R. and Christie, John M.},
doi = {10.1016/S1360-1385(02)02245-8},
issn = {13601385},
journal = {Trends in Plant Science},
number = {5},
pages = {204--210},
pmid = {11992825},
title = {{Phototropins 1 and 2: Versatile plant blue-light receptors}},
volume = {7},
year = {2002}
}

@article{Luesse2010,
author = {Luesse, Darron R. and Deblasio, Stacy L. and Hangarter, Roger P.},
doi = {10.1093/jxb/erq242},
issn = {00220957},
journal = {Journal of Experimental Botany},
number = {15},
pages = {4387--4397},
pmid = {20693413},
title = {{Integration of phot1, phot2, and PhyB signalling in light-induced chloroplast movements}},
volume = {61},
year = {2010}
}

@article{Labuz2022,
author = {{\L}abuz, Justyna and Sztatelman, Olga and Hermanowicz, Pawe{\l}},
doi = {10.1093/jxb/erac271},
issn = {14602431},
journal = {Journal of Experimental Botany},
number = {18},
pages = {6034--6051},
title = {{Molecular insights into the phototropin control of chloroplast movements}},
volume = {73},
year = {2022}
}

@software{FIJI,
	author = {Schindelin, Johannes and Arganda-Carreras, Ignacio and Frise, Erwin and Kaynig, Verena and Longair, Mark and Pietzsch, Tobias and Preibisch, Stephan and Rueden, Curtis and Saalfeld, Stephan and Schmid, Benjamin and Tinevez, Jean-Yves and White, Daniel James and Hartenstein, Volker and Eliceiri, Kevin and Tomancak, Pavel and Cardona, Albert},
	doi = {10.1038/nmeth.2019},
	id = {Schindelin2012},
	isbn = {1548-7105},
	journal = {Nature Methods},
	number = {7},
	pages = {676--682},
	title = {Fiji: an open-source platform for biological-image analysis},
	url = {https://doi.org/10.1038/nmeth.2019},
	volume = {9},
	year = {2012},
    month = {Jul}}

@software{trackpy,
  author       = {Allan, Daniel B. and
                  Caswell, Thomas and
                  Keim, Nathan C. and
                  van der Wel, Casper M. and
                  Verweij, Ruben W.},
  title        = {soft-matter/trackpy: Trackpy v0.5.0},
  month        = apr,
  year         = 2021,
  publisher    = {Zenodo},
  version      = {v0.5.0},
  doi          = {10.5281/zenodo.4682814},
  url          = {https://doi.org/10.5281/zenodo.4682814}
}

\end{document}